\documentclass[preprint,12pt]{elsarticle}

\usepackage{graphicx}
\usepackage{amssymb}

\usepackage{lineno}

\journal{NIMA}

\begin{document}

\begin{frontmatter}


\title{ Resistive AC-Coupled Silicon Detectors: principles of operation and first results from a combined analysis of beam test and laser data}



\author{M. Tornago\footnote{email: marta.tornago@edu.unito.it} $^{a,b}$, R. Arcidiacono $^{b,c}$, N. Cartiglia $^{b}$, M. Costa $^{a,b}$,\\ M. Ferrero $^{b,c}$, M. Mandurrino $^{b}$, F. Siviero $^{a,b}$, V. Sola $^{b}$, A. Staiano $^{b}$,\\
A. Apresyan $^{d}$, K. Di Petrillo $^{d}$, R. Heller $^{d}$, S. Los $^{d}$,\\ G. Borghi $^{f,g}$, M. Boscardin $^{f,g}$, G-F Dalla Betta $^{e,f}$, F. Ficorella $^{f,g}$,\\ L. Pancheri $^{e,f}$, G. Paternoster $^{f,g}$\\ H. Sadrozinski$^{h,i}$, A. Seiden$^{h,i}$ }

\address{$^{(a)}$Universit\'a degli Studi di Torino, Italy\\$^{(b)}$INFN Torino, Italy\\$^{(c)}$Universit\'a degli Studi del Piemonte Orientale\\$^{(d)}$Fermi National Accelerator Laboratory, Batavia, USA\\$^{(e)}$Universit\'a degli Studi di Trento, Italy\\$^{(f)}$INFN-TIFPA, Trento, Italy\\$^{(g)}$Fondazione Bruno Kessler, Trento, Italy\\$^{(h)}$University of California, Santa Cruz\\$^{(i)}$Santa Cruz Institute for Particle Physics}

\begin{abstract}
This paper presents the principles of operation of Resistive AC-Coupled Silicon Detectors (RSDs) and measurements of the temporal and spatial resolutions using a combined analysis of laser and beam test data.
RSDs are a new type of n-in-p silicon sensor based on the Low-Gain Avalanche Diode (LGAD) technology, where the $n^+$ implant has been designed to be resistive, and the read-out is obtained via AC-coupling. 
The truly innovative feature of RSD is that the signal generated by an impinging particle is shared isotropically among multiple read-out pads without the need for floating electrodes or an external magnetic field. 
Careful tuning of the coupling oxide thickness and the $n^+$ doping profile is at the basis of the successful functioning of this device.
Several RSD matrices with different pad width-pitch geometries have been extensively tested with a laser setup in the Laboratory for Innovative Silicon Sensors in Torino, while a smaller set of devices have been tested at the Fermilab Test Beam Facility with a 120 GeV/c proton beam. The measured spatial resolution ranges between $2.5\; \mu m$ for 70-100 pad-pitch geometry and $17\; \mu m$ with 200-500 matrices, a factor of 10 better than what is achievable in binary read-out ($bin\; size/ \sqrt{12}$). Beam test data show a temporal resolution of $\sim 40\; ps$ for 200-$\mu m$ pitch devices, in line with the best performances of LGAD sensors at the same gain.
\end{abstract} 


\end{frontmatter}


\section{Introduction}
\label{S:1}

\begin{figure}[th!]
    \centering
    \includegraphics[scale=0.35]{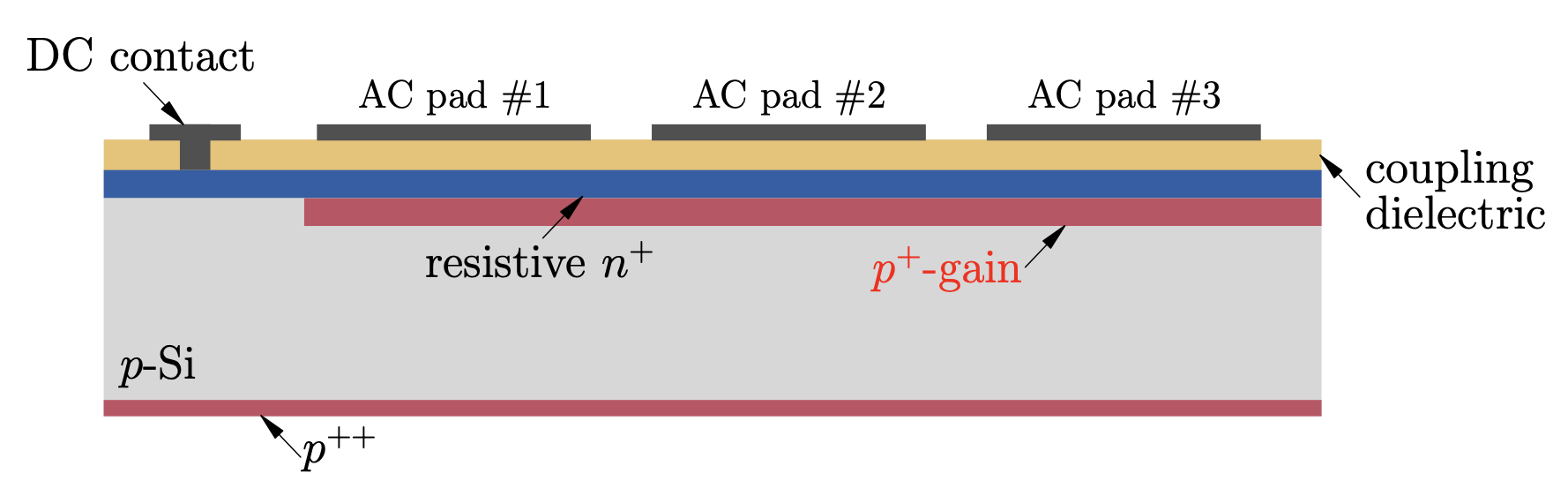}
    \caption{Cross-section of RSDs internal structure: their properties are based on the combination of a resistive $n^+$ layer and a coupling dielectric oxide, allowing a local AC-coupling.}
    \label{fig:RSD}
\end{figure}

AC-Coupled Low Gain Avalance Diodes (AC-LGADs) \cite{NC15, BNL} are a new generation of silicon devices optimized for high-precision 4D tracking and conceived for experiments at future colliders. They are n-in-p sensors based on the LGAD technology with two additional key features (Figure \ref{fig:RSD}): the AC-coupling of the read-out, occurring through a dielectric layer, and a continuous resistive $n^+$ implant. Given the presence of the resistive $n^+$ layer, AC-LGADs are called Resistive Silicon Detectors (RSD). RSD devices are provided with one continuous gain layer, and the read-out segmentation is obtained simply by the position of the AC pads; therefore, this design allows to reach $100\%$ fill-factor. 

The remarkable feature of this design is that it leads naturally to signal sharing among pads. 
Internal signal sharing, in combination with internal gain, opens a new avenue for high precision tracking without relying only on shrinking the pitch: a position resolution of $\sim 5\; \mu m$ is achieved with 200-$\mu m$ pitch combining the signals from several pads.
Instead of focusing on how to design the smallest possible pixel, the RSD design focuses on how to maximize the uniformity of signal sharing and to find the right balance between signal sharing, gain, noise, and reconstruction. 

The first RSD production (RSD1) \cite{MM1,MM2} has been manufactured in 2019 by the Fondazione Bruno Kessler. The batch includes 15 wafers, 11 Si-on-Si and 4 epitaxial, with varying $n^+$, gain and p-stop doping profiles, and different oxide thicknesses. Each wafer contains several devices with different geometries; the ones used for the measurements presented in this paper are square matrix sensors with a varying number of pads, pitch, and pad size, but equal oxide thickness and $n^+$ doping profile.

\section{RSD signal properties}

Signal formation in RSDs happens in three phases \cite{NC1}.
\begin{itemize}

\item[(i)] In the first one, the drifting e/h pairs produce a direct charge induction on the $n^+$ conductive implant, similar to a standard LGAD. This initial signal is not seen on the AC-metal pads unless the metal pad is directly above the impact point. 
\item[(ii)] The signal spreads toward ground on the $n^+$  layer. The AC metal pads, grounded via the read-out electronics, offer to fast signals a path to ground with a resistance far lower than that provided by the $n^+$  resistive sheet. For this reason, the fast signal (about 1 ns for 50-$\mu$m thick sensors) becomes visible on the AC pads, charging the capacitors formed by the AC metal pads and the n+ layer. The signal in the AC pads is seen with a delay that increases with distance from the impact point. If the metal pad is right above the impact point, there is no delay.  Signal sharing in AC-LGADs happens on the surface of the $n^+$  layer and does not require long drift lines. This mechanism enables to combine charge sharing and thin detectors. This second phase generates the first lobe of the signal, identical to the one created in an equivalent DC LGAD, as shown in Figure \ref{fig:signal}. The effect of trapped charges at the Si-SiO2 interface does not spoil this mechanism in non-irradiated sensors. The effect of irradiation will be evaluated in a subsequent publication. 

\begin{figure}[th!]
    \centering
    \includegraphics[scale=0.25]{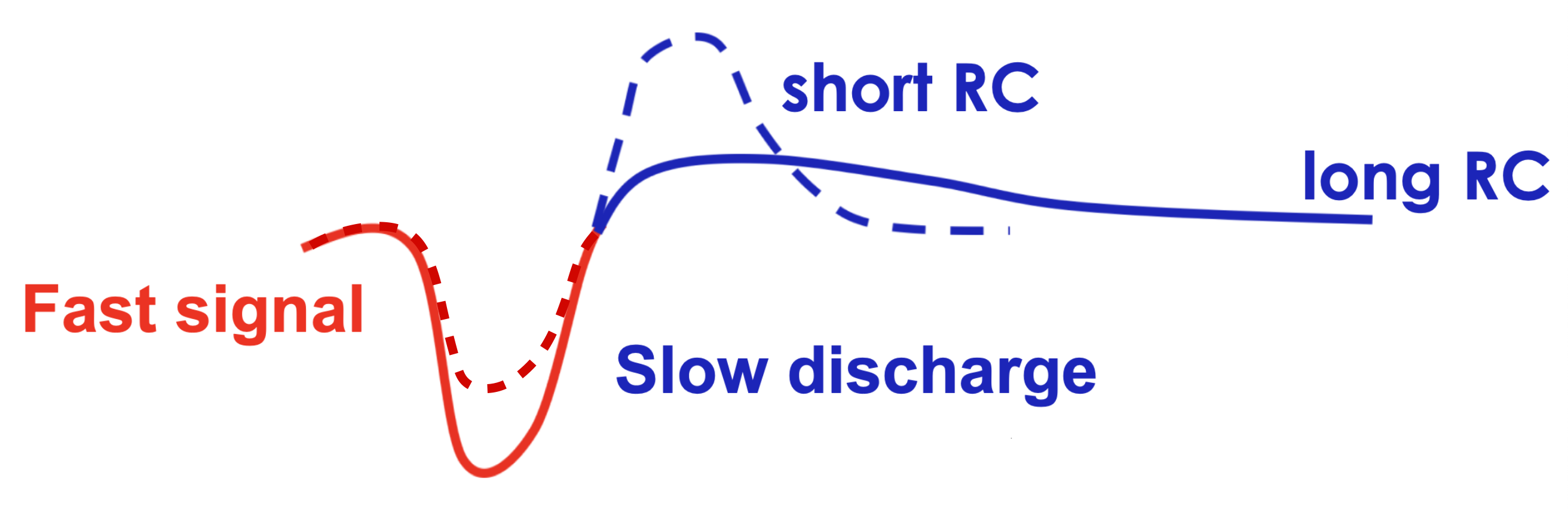}
    \caption{A typical signal generated by an RSD device, characterized by a first fast negative lobe, identical to the signal of an equivalent DC LGAD, and a slow positive lobe by which the AC pad discharges.}
    \label{fig:signal}
\end{figure}

\item[(iii)] In the last step, the AC pads discharge, generating the second lobe of the signal, with opposite polarity with respect to the first one. The shape of this lobe depends on the resistance to ground $R$ and the pad capacitance $C$, the $RC$ time constant. Systems with a small $RC$ will have signals with a larger and shorter positive lobe since they need to discharge the same amount of charge in a shorter time.
The value of the $RC$ time constant also affects the first lobe: if the $RC$ is too short, the first lobe will be smaller due to ballistic deficit. 

\end{itemize}

\begin{figure}[th!]
    \centering
    \includegraphics[scale=0.15]{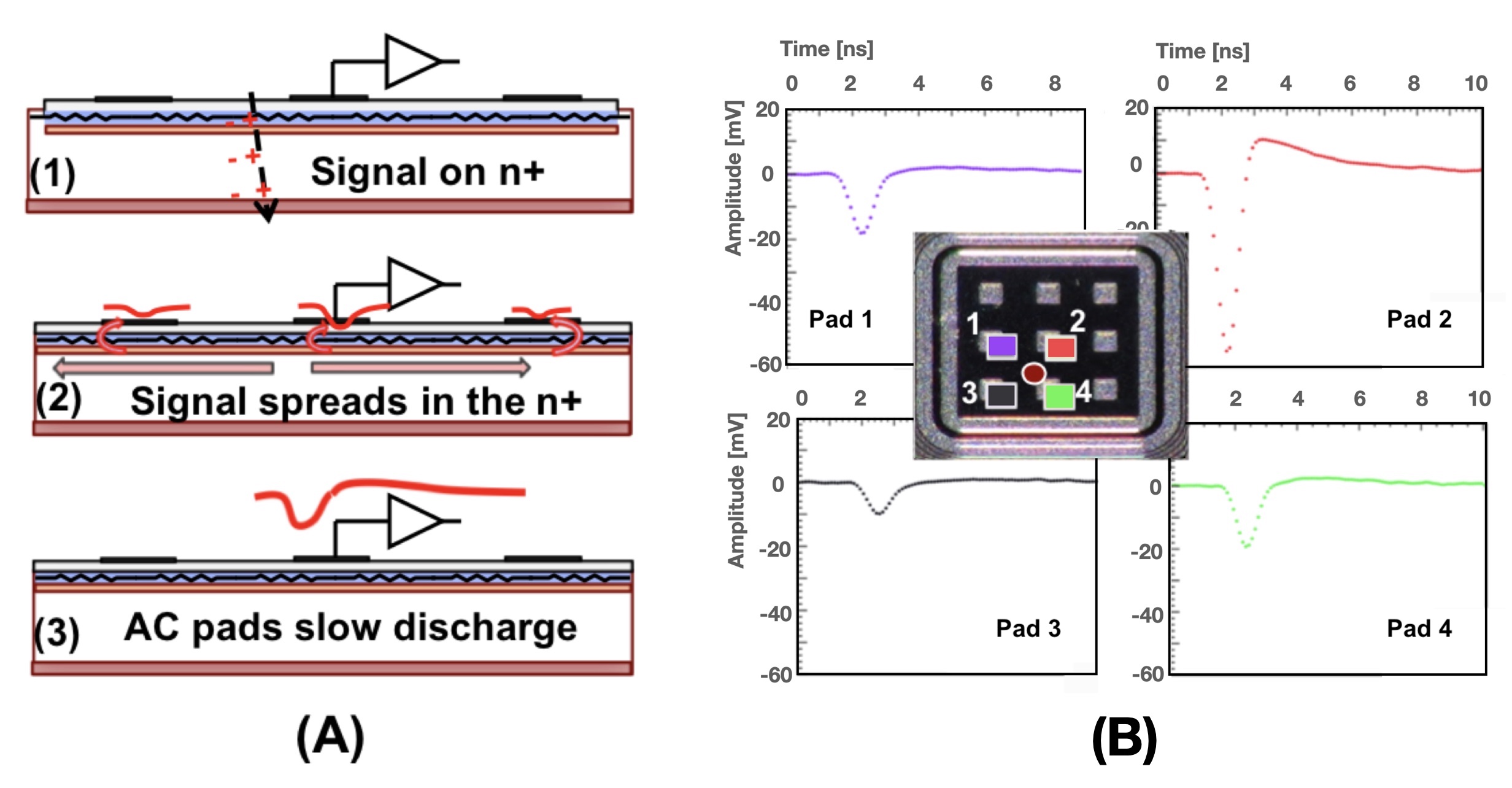}
    \caption{Sketch of the three phases of RSD signals formation (A); example of signal sharing (B): four AC pads see a signal for a hit position corresponding to the red point on the RSD picture.}
    \label{fig:signalshare}
\end{figure}

Figure \ref{fig:signalshare} shows a sketch describing the signal formation phases (A), and the signals seen on the 4 pads surrounding the laser hit located in the red spot (B). 

\section{RSD Reconstruction Models}
\label{S:2}

Extensive studies with a laser setup have been performed in the Laboratory for Innovative Silicon Sensors in Torino, using RSD matrices with varying pitch and pad size. The main objectives of these measurements were the study of the signal formation and the measurement of the spatial and temporal resolutions as a function of the RSD pitch and pad size.

\subsection{Signal sharing: the Logarithmic Attenuation Model}
 
\begin{figure}[th!]
    \centering
    \includegraphics[scale=0.15]{ 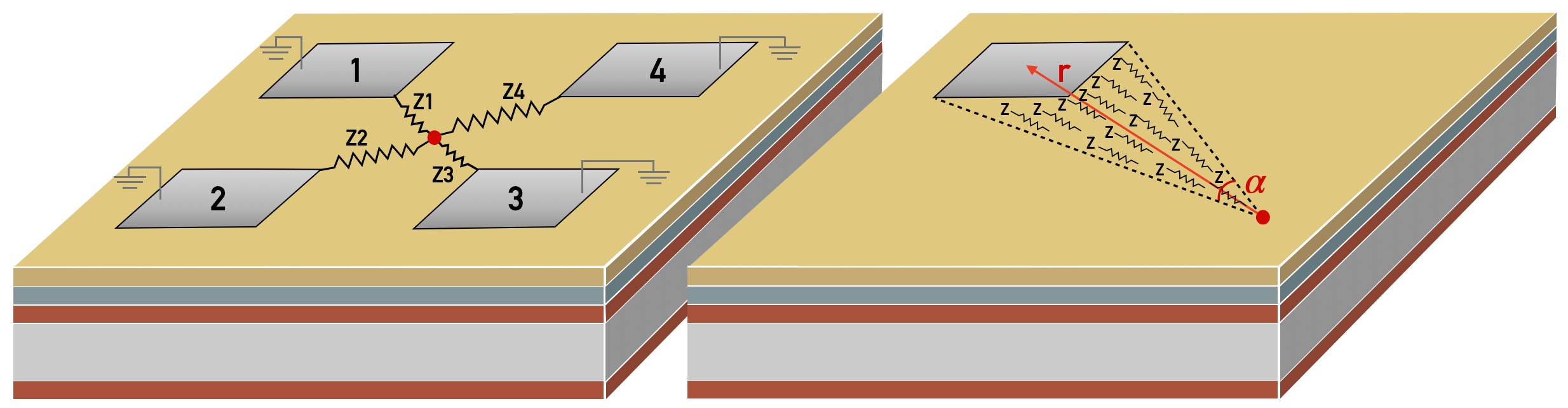}
    \caption{Sketch of an RSD, with the resistance seen by a signal while propagating.}
    \label{fig:master}
\end{figure}

In RSDs, a hit generates signals on multiple pads.
This can be explained by comparing an RSD to a current divider. In such divider, 
the current  in a given branch is:
\begin{equation}
\label{eq:divider}
    I_i = I_0 \frac{\frac{1}{R_i}}{\sum_1^n \frac{1}{R_i}},
\end{equation}

where $I_0$ is the total current, $I_i$ is the current in the branch $i$ and $R_i$ the resistance. In an RSD, the situation is very similar (see  Figure \ref{fig:master} - left), with the branches of the divider being the resistances of the triangular surface connecting the impact point to each of the neighboring pads. As the signal travels from the hit point to the pad, it spreads over a larger area: the resistance $R_i$ seen by the signal does not scale linearly with the distance (Figure \ref{fig:master} right). Specifically, the resistance per unit length decreases as a function of the distance $r$ as: 
\begin{equation}
    R(r) \propto \rho \frac{dr}{r \alpha}.
\end{equation}

where $\rho$ is the $n^+$ resistivity expressed in $\Omega/m$ and $\alpha$ is the angle of view of the pad.

Therefore, the expression for the resistance seen by a signal propagating towards a pad at a distance $d$ is:
\begin{equation}
    R(d) \propto \frac{\rho}{\alpha}\int_{d_0}^d \frac{1}{r} dr \propto \rho \frac{ln(d/d_0)}{\alpha}.
    \label{eq:resistance}
\end{equation}

The lower limit of the integral is the transverse size of the hit. Considering the lateral extent of the delta rays and ionization, a distance of 1 micron is used, $d_0 = 1\; \mu m$.  For the upper limit $d$, the physical cutoff is given by a distance comparable with the pitch size. In our experimental setup, we have checked the validity of this formula up to distances of about  400-500 $\mu$m.  

Combining Equation (\ref{eq:divider}) with Equation (\ref{eq:resistance}), it is possible to calculate how a signal is shared among pads: 
\begin{equation}
    F_i(\alpha_i,d_i)= \frac{\frac{\alpha_i}{ln(d_i/{d_0})}}{\sum_1^n \frac{\alpha_i}{ln(d_i/{d_0})}},
        \label{eq:masterformula}
\end{equation}
where $F_i$ is the fraction of the total signal amplitude seen on the pad $i$,  $d_i$ the distance from the hit point to the pad $i$ metal edge, and $\alpha_i$ the pad $i$ angle of view.  Equation (\ref{eq:masterformula}) predicts without any free parameter how a signal is shared among pads for every RSD geometry, $n^+$ resistivity, and dielectric thickness, as the signal sharing depends on the relative resistance of each path, and not on its absolute value.

This formulation allows to highlight some properties of signals in RSDs:
\begin{itemize}
    \item the signal seen by a pad depends on how many and how close the other pads are,
    \item the number of pads that record a signal depends on the hit location,
    \item if the hit is located on a metal pad, and the metal size is large enough (more than about 15x15 $\mu m^2$), the signal is almost completely absorbed by a single pad since the path impedance from the hit point to this pad is zero,
    \item a floating pad does not contribute to charge sharing, as the impedance to ground is infinite,
    \item the sum of all the recorded signals is constant.
\end{itemize}

Since the $n^+$ layer is resistive, there is a delay between the hit time and the signal formed on a given pad. This delay is proportional to the value of the impedance and of the system capacitance.  The signal arrival time on a pad at a distance $d$, and with an angle of view $\alpha$ from the hit point, is: 
\begin{equation}
    t(d,\alpha) = t_0 + \gamma \frac{ln(d/{d_0})}{\alpha},
    \label{eq:time}
\end{equation}
where $t_0$ is the hit time and $\gamma$ is the delay factor that can be extracted from experimental data.\\
Equations (\ref{eq:masterformula}) and (\ref{eq:time}) are called the RSD main formulas: they allow to calculate the fraction of the signal seen on each pad, and its delay. 

\subsection{Signal sharing: the Linear Attenuation Model}
A simple model of signal attenuation with distance, the \textit{linear attenuation} model (LinA), was also used.  The model assumes that the signal on a pad decreases linearly with the distance from the impact point and increases linearly with the angle of view \cite{NC2}. In the LinA model, the fraction of the total signal seen by a pad is given by: 
\begin{equation}
    F_i(d_i,\alpha_i) = \frac{[1 - \beta * d_i]*\alpha_i}{\sum_i^n [1 - \beta * d_i]*\alpha_i},
    \label{eq:LA}
\end{equation}
where $d_i$ is the distance from the pad metal edge, $\alpha_i$ the angle of view, and $\beta$ is the attenuation factor. 
The strength of this model is the presence of a tunable parameter, $\beta$. For each geometry, $\beta$ is determined from data as the value that minimises the position resolution. 

\begin{figure}[th]
    \centering
    \includegraphics[scale=0.45]{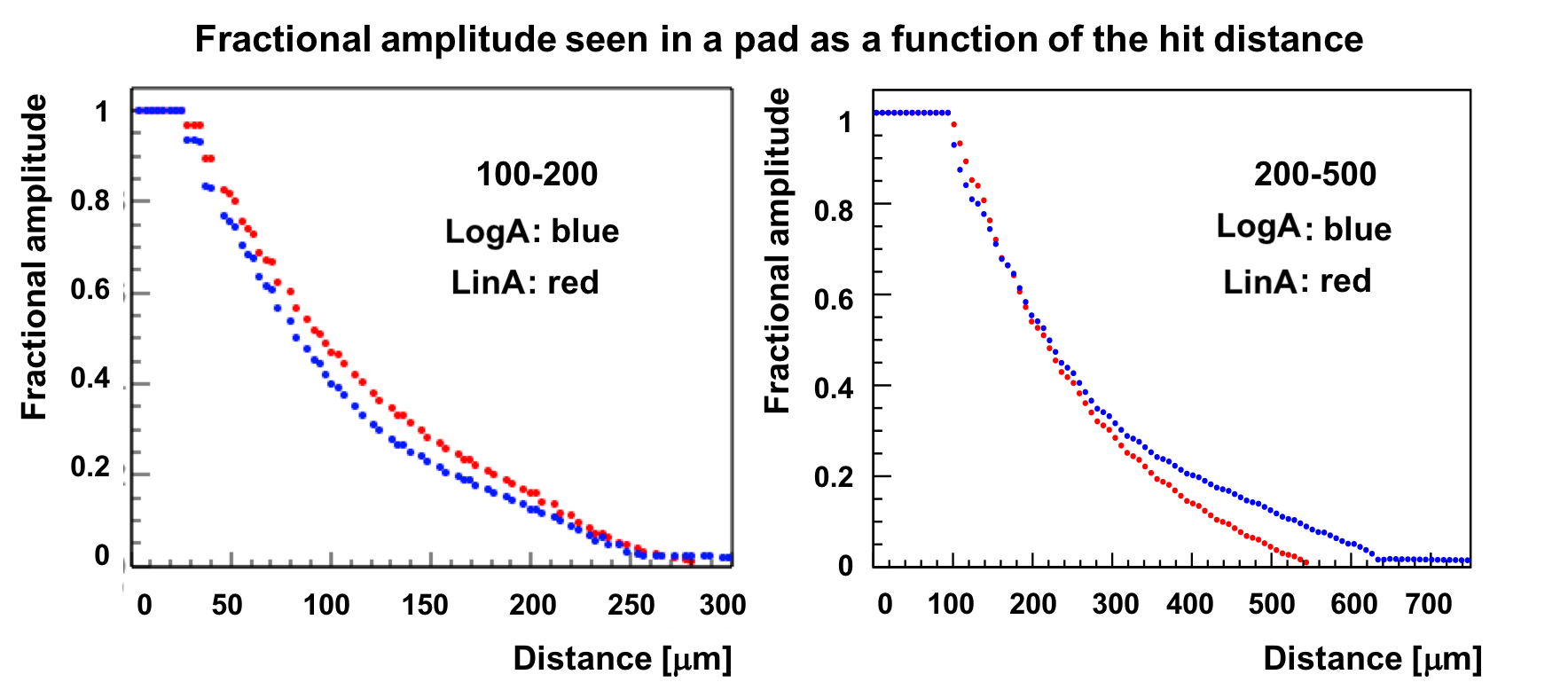}
    \caption{Fraction of the total signal as a function of the hit distance from the center of the AC pads, for the LogA (blue) and the LinA model (red), plotted for the pad-pitch geometries 100-200 ($\beta = 0.003 \mu m^{-1}$ )and 200-500 ($\beta = 0.002 \mu m^{-1}$ ).}
    \label{fig:1Dampvsdist}
\end{figure}

The LinA model Equation for the time delay also has a linear dependence on the hit distance:
\begin{equation}
\label{eq:lindelay}
    t(d) = t_0 + \zeta*d,
\end{equation}
where $\zeta$ is the delay factor extracted from data.\\
Figure \ref{fig:1Dampvsdist} illustrates an example of the predictions from the two models. The plot shows the fraction of the amplitude seen in a pad as a function of its distance from the hit position as predicted by the LogA (in blue) and by the LinA model (in red) for the 100-200 and 200-500 pad-pitch geometries. The LogA and the LinA-tuned predictions are very similar for the smaller geometry, while they differ at large distances in the 200-500 case. 

\subsection{Position reconstruction using the LogA and LinA methods}
\label{S:reco}
\begin{figure}[thb]
    \centering
    \includegraphics[scale=0.13]{ 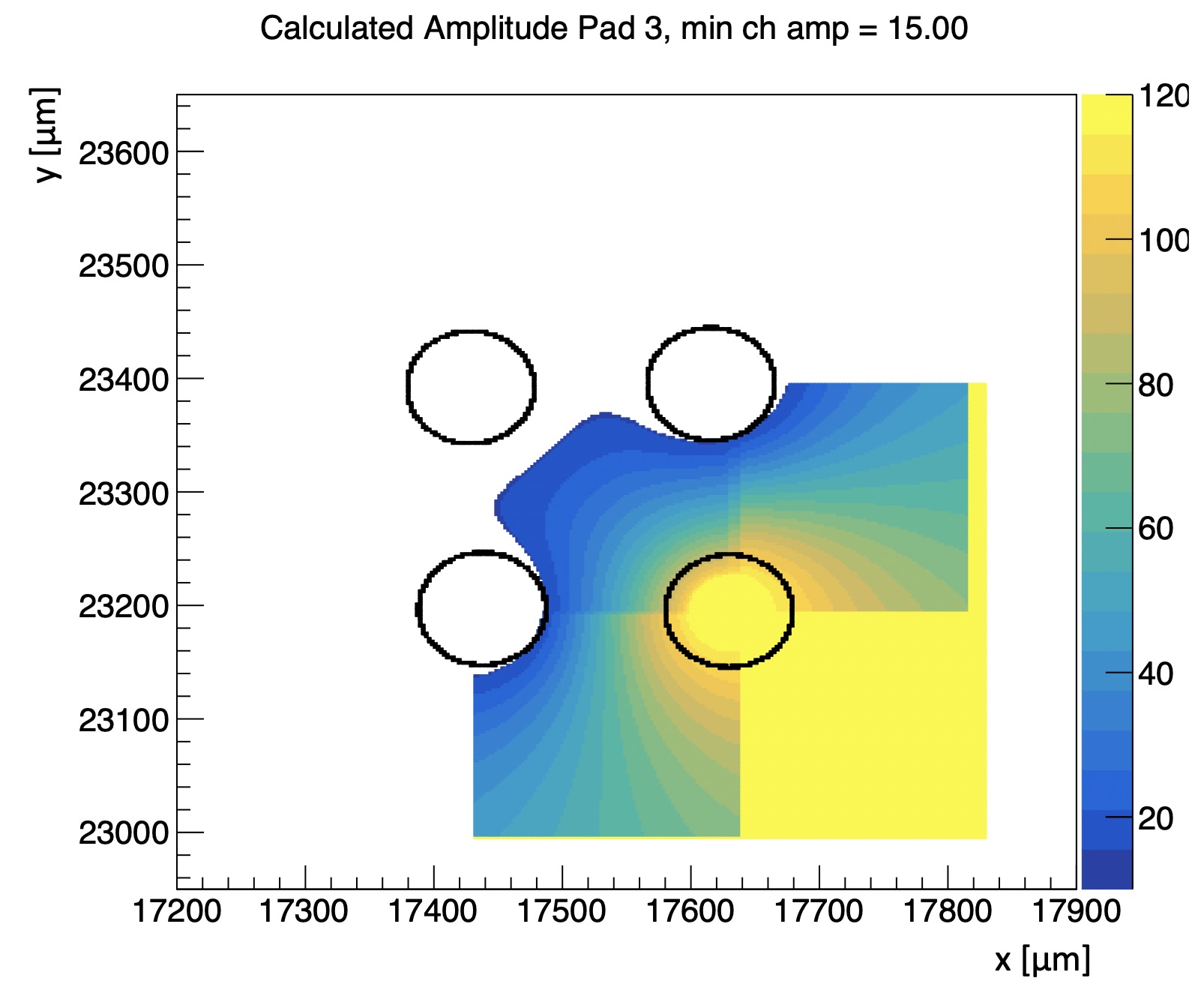}
    \caption{Amplitude seen by one AC pad (bottom-right) as a function of the 2D hit position on the RSD surface, calculated with the RSD LogA formulas, assuming a total amplitude of 120 mV.}
    \label{fig:lut}
\end{figure}

A point on the RSD surface is uniquely identified by the relative amplitudes seen by the nearby pads. Exploiting this remarkable property, the location of the hit can be reconstructed very accurately. 
Figure \ref{fig:lut} shows the amplitude seen in a given pad as a function of the hit position on the RSD surface, according to the RSD LogA model (assuming a total amplitude of 120 mV). 

The reconstruction of the hit position is achieved with the following procedure:
\begin{itemize}
    \item The total signal $A_{tot}$ amplitude is calculated as the sum of the amplitudes $A[i]$ seen in all active pads, $A_{tot}=\sum_{i} A[i]$, with $A[i]$ larger than a given threshold, chosen to exclude the noise;
    \item The fraction of the total amplitude seen in each pad is calculated as $(A[i]/A_{tot})_{Meas}$; 
    \item The set of fractions is compared with the fractions predicted in each x-y bin (using either LogA or LinA). The hit position is the x-y bin that minimizes the following chi-square:
    \begin{equation}
        \chi^2=\sum_{i}\left[\left(\frac{A[i]}{A_{tot}}\right)_{Meas}-  \left(\frac{A[i]}{A_{tot}}\right)_{Calc}\right]^2;
    \end{equation}
    \item The accuracy can be increased by performing a local interpolation around the minimum.
\end{itemize}


\subsection{RSD as a Discretized Positioning Circuit}
\begin{figure}[th!]
    \centering
    \includegraphics[scale=0.4]{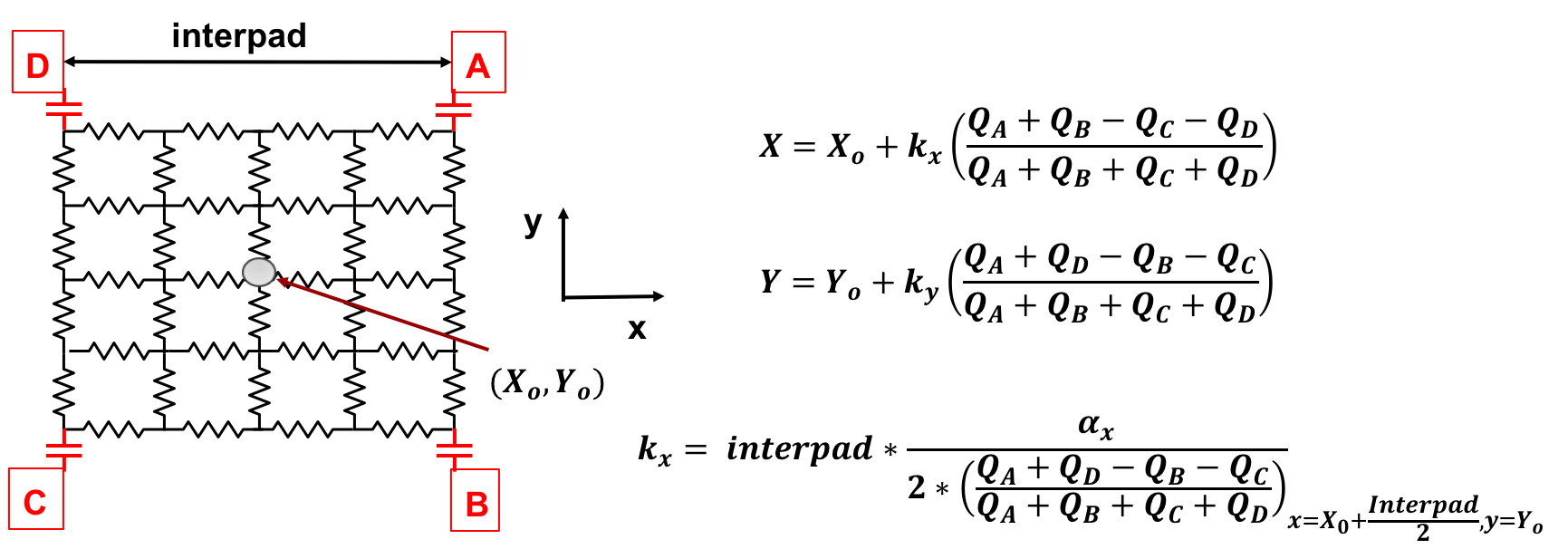}
    \caption{Left side: the DPC representation of an RSD. Right side: the $x$ and $y$ expressions. $X_o,Y_o$ are the coordinates of the central point, $Q_{A,B,C,D}$ the charge collected on each of the pad, $k$ the charge imbalance along $x$  (or $y$) measured at the $x$ coordinate of the pads with the higher $x$ coordinate. }
    \label{fig:DPC_NIMA}
\end{figure}

In an RSD, each group of four pads delimiting a square can be considered the four corners of a  Discretized Positioning Circuit (DPC). This is shown on the left side of Figure \ref{fig:DPC_NIMA}.

This type of network is commonly used in positron emission tomography detectors to read out large arrays of  Silicon Photomultiplier (SiPM) with only four read-out amplifiers. 
In DPC, a SiPM is positioned in each internal node, and an amplifier \cite{Park_2017} is positioned at each corner.  
The $x$ and $y$ positions of the hit are determined by exploiting the charge imbalance between pairs of pads along the $x$ and $y$ direction. The application of the DCP read-out scheme to RSD is shown on the right side of Figure \ref{fig:DPC_NIMA}. The coefficient $k_{x,y}$ is the product of 3 terms: the length of the region where the measurement is performed (the interpad region), the inverse of the maximum one-side charge imbalance, and a geometrical factor $\alpha_{x,y}$.

The strong point of the DPC approach is that the $x$ and $y$ positions are determined without any prior assumption on the sharing law. It requires only the measurement of the charge on the 4 closest pads to the hit. 

\subsection{Design optimization}
As seen in the previous paragraph, signal sharing is germane to the RSD design and depends upon the pad shape and positioning. In order to obtain a detector with a very good and uniform spatial resolution, the position and shape of the pads should be optimized. 

\begin{figure}[th!]
    \centering
    \includegraphics[scale=0.15]{ 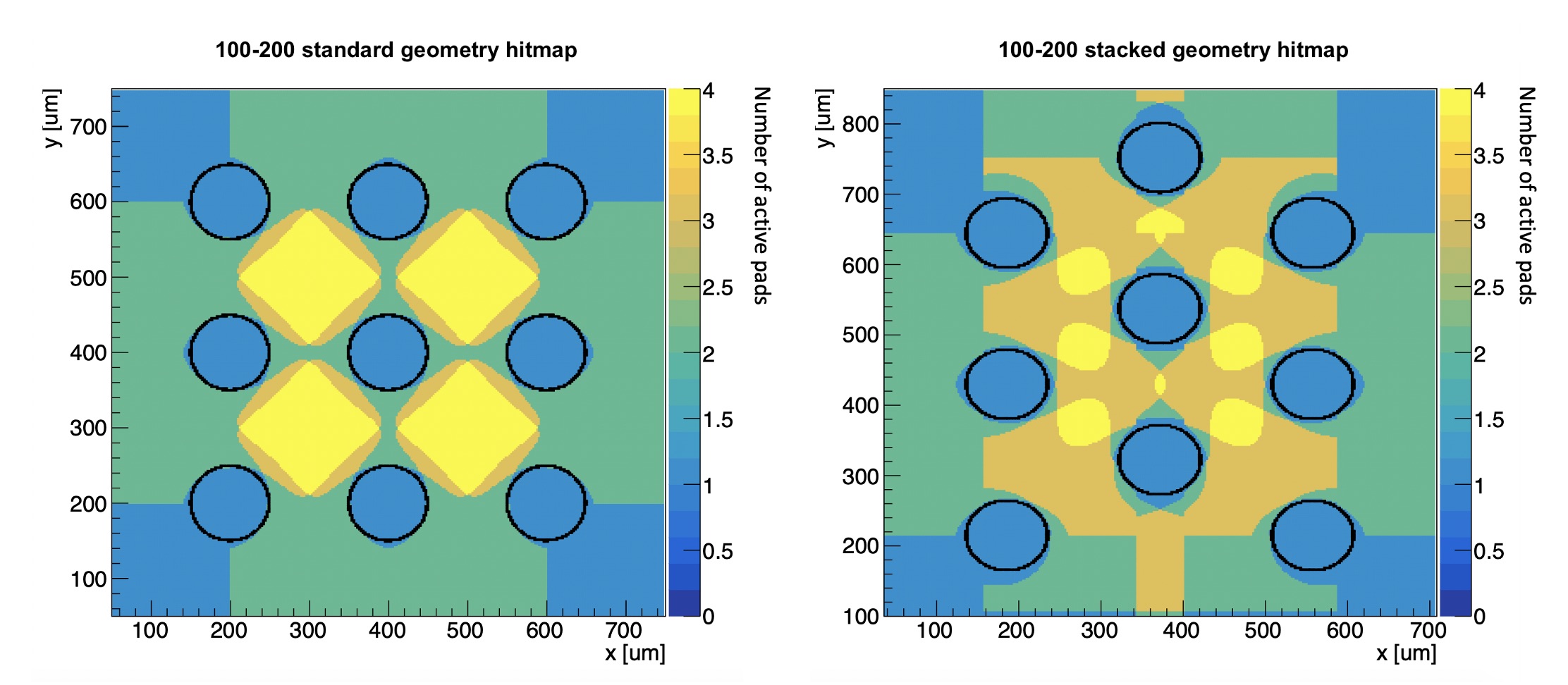}
    \caption{2D maps representing how many pads see a signal depending on the hit position in a standard $3\times 3$ 100-200 pad-pitch matrix (left) and in a device with a modified pad layout (right). Black circles identify the edge of the metal pads, here considered as circles for simplicity.}
    \label{fig:tappeto}
\end{figure}


Figure \ref{fig:tappeto} (left) shows the number of pads reading a signal above noise, for a $3\times 3$ 100-200  matrix, with pads positioned on a squared grid. The orange and yellow areas indicate the region where hits are reconstructed with three and four pads respectively: information from at least three pads allows to uniquely identify the hit position. This geometry shows two shortcomings: 
\begin{itemize}
    \item  the reconstruction with three pads is less accurate than that with four since the hit reconstruction is biased  closer to the three active pads
    \item the green area, where only two pads record a signal, and it is no longer possible to determine the hit position, is not negligible. 
\end{itemize}

An alternative design is shown in Figure \ref{fig:tappeto} (right), where the AC pads are placed in a triangular geometry. This geometry has two advantages: the signal in each pad is higher since the split is mostly among 3 pads and not four, and the area where only two pads see a signal is very small, reaching almost full efficiency.

\subsubsection{Design of the metal pad}

\begin{figure}[th!]
    \centering
    \includegraphics[scale=0.4]{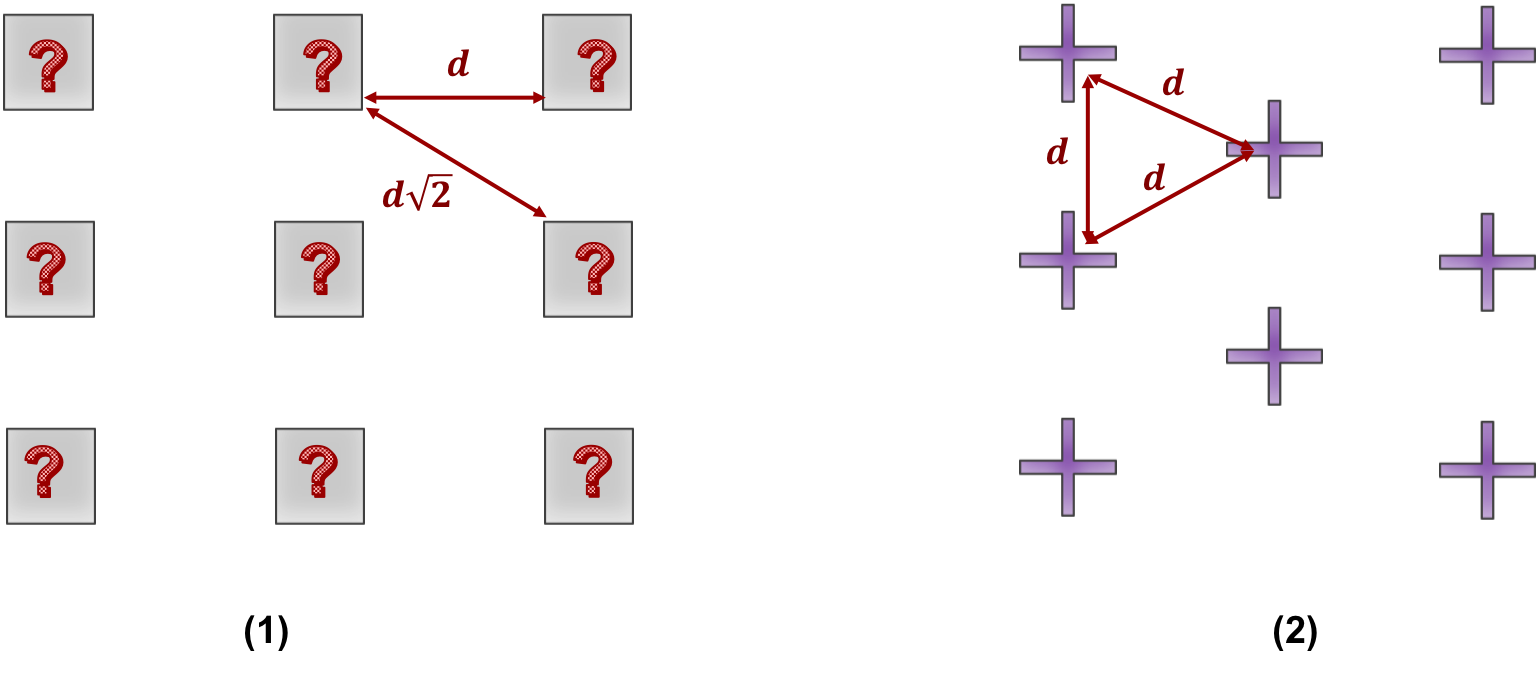}
    \caption{Present metal pads design in a $3\times3$ RSD matrix (1) and a possible optimization of the pads shape and layout (2)}
    \label{fig:read_out_position}
\end{figure}

In the designs simulated in Figure \ref{fig:tappeto} there is no signal sharing when the particle hits a metal pad. This has been measured experimentally, shooting the laser signal in a narrow slit without metal, placed inside the metal pad. Moving along the slit, from the center to the periphery of the pad, the signal sharing starts $\sim 10-20\; \mu m$ from the metal edge.  With this information, it is possible to redesign the metal pad maximizing the signal sharing. The new design should include two important aspects:
\begin{itemize}
    \item the angle of view in Equation (\ref{eq:masterformula}) should not be too small,
    \item the capacitance of the pad should be large enough to prevent a ballistic deficit. 
\end{itemize}
Figure \ref{fig:read_out_position} proposes an optimization of the metal pad disposition: (i) the angle of view remains large, (ii) the pad capacitance is enough to prevent ballistic deficit, and (iii) the metal width of the arms does not prevent signal sharing. 
This design will be investigated in future productions.

\section{Measurements of the RSD properties using a  TCT laser setup}

The properties of the RSD sensors have been studied using the Transient Current Technique (TCT), which exploits the signal induced in the sensor by a laser. The TCT system employed in this study, manufactured by Particulars, is equipped with an IR picosecond laser, with a minimum spot size of $\sim 10\; \mu m$, and with a micro-metrical x-y moving stage. The intensity of the laser was set to reproduce the amount of ionization typical of a MIP in a sensor of the same active thickness, about 0.5 fC. 

\begin{figure}[th!]
    \centering
    \includegraphics[scale=0.17]{ 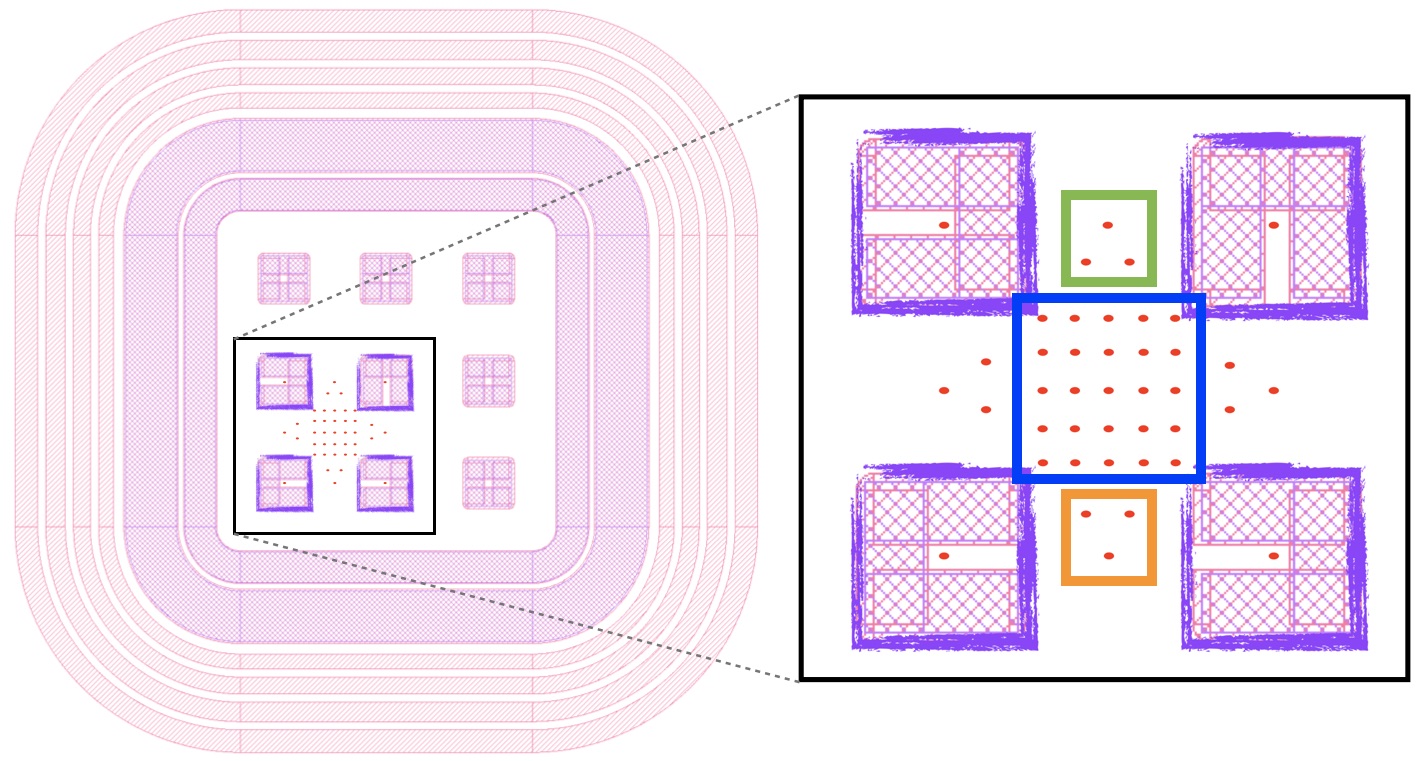}
    \caption{Sketch of a 100-200 pad-pitch matrix measured in the TCT set-up. The red dots represents the locations of the laser shots.}
    \label{fig:laser}
\end{figure}

The RSD1 production, as previously mentioned, comprises 15 wafers that differ for the values of the $n^+$ resistivity and coupling capacitance. All measurements have been obtained with the wafers with the highest resistivity and value of the coupling capacitance in order to minimize the ballistic deficit. Wafers with higher ballistic deficits have smaller signals that yield to worse performances. 
Several RSD matrices, with different pitch-pad sizes, have been characterized with the TCT setup. Each sensor has been tested at bias voltages corresponding to values of gain between 8 and 25. For each matrix, the laser is shot in multiple points in the area among the pads, as shown in Figure \ref{fig:laser}. The four AC pads closest to the hit points have been read out simultaneously and their signals recorded using a 40Gs/s, 4 GHz BW oscilloscope ($\sim500$ waveforms per position \cite{RA, GP, MT}). For each laser shot, the hit position is reconstructed using the method described in section \ref{S:reco}. The spatial resolution at each position is evaluated as the sigma of the difference between the laser shot coordinates and the reconstructed position $(x,y)_{reco}-(x,y)_{laser}$. 

\begin{figure}[th!]
    \centering
    \includegraphics[scale=0.12]{ 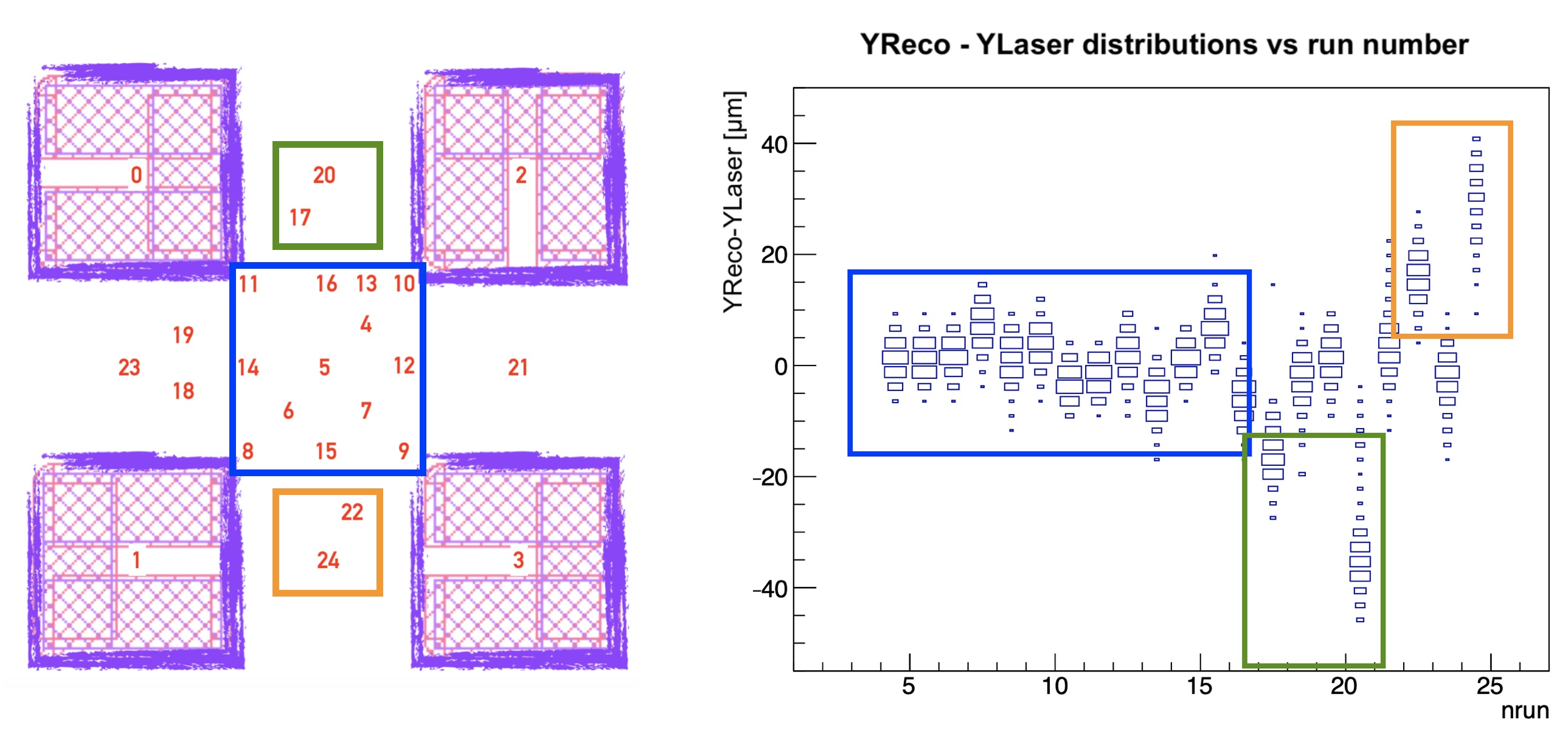}
    \caption{$y_{reco}-y_{laser}$ distributions plotted for single laser shots in different regions of a 100-200 pad-pitch matrix.}
    \label{fig:spaceresvsregion}
\end{figure}

The left side of Figure \ref{fig:spaceresvsregion} indicates with numbers the positions where the laser was shot during a given acquisition sequence. On the right side, the spatial resolution of the y coordinate is reported as a function of the acquisition number.

 Points between two pads, like those in the green and orange squares, are poorly reconstructed since for these areas the signal is recorded only by two pads. The $(x,y)_{reco}- (x,y)_{laser}$ distributions for these positions are wider and shifted. 

Signals shot inside the metal opening of the pads suffer from even greater uncertainties, as the only information is the amplitude recorded by the only active pad. For this type of events, the spatial resolution corresponds to the case of binary read-out, $pad\;size/\sqrt{12}$. 
Only the shots with a signal completely contained in the four read-out AC pads are considered in the following studies (see the blue square in Figure \ref{fig:spaceresvsregion}). 

\subsection{Study of signal sharing among pads}

\begin{figure}[th!]                                      
    \centering
    \includegraphics[scale=0.75]{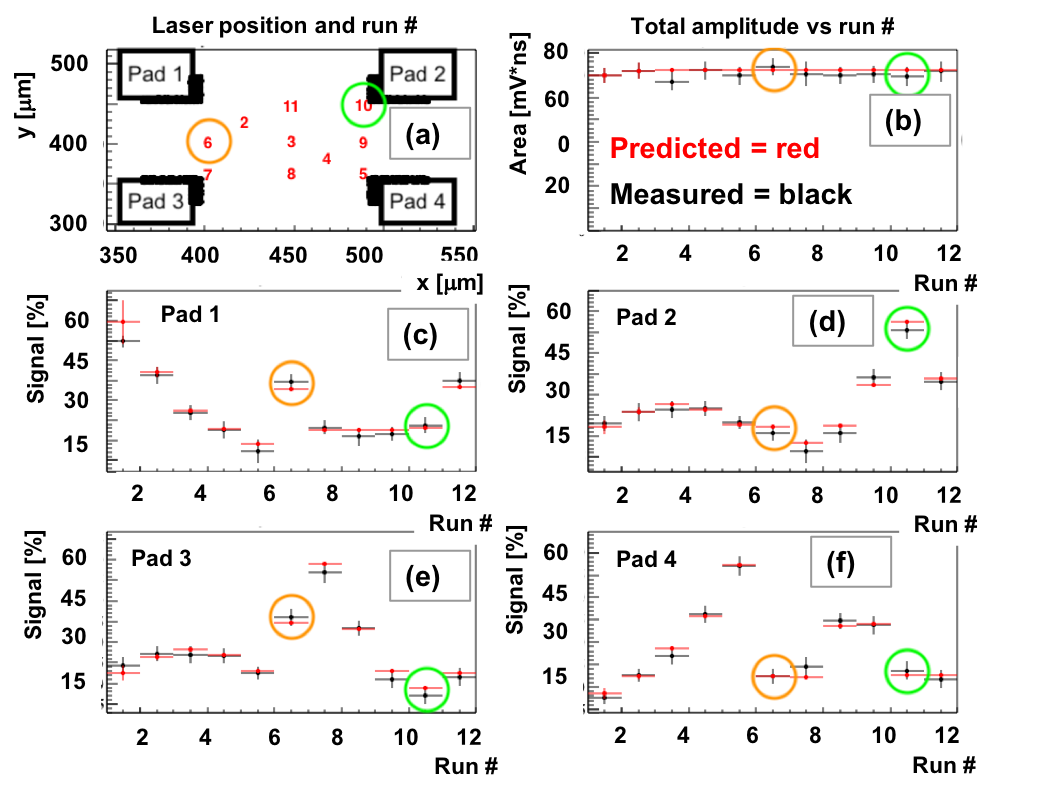}
    \caption{Map with positions of the laser runs for the 100-200 pad-pitch geometry (a); the sum of the signal amplitudes of the four read-out pad for each run (b). Plots (c)-(f) show the percentage amplitude seen by each AC pad as a function of the laser run number. Black points represent the predicted values obtained with the LogA formula, while the red ones are the measured values.}
    \label{fig:laser_signalsharing}
\end{figure}

TCT studies have also been used to verify the predictions of Equation (\ref{eq:masterformula}). 
Figure \ref{fig:laser_signalsharing} (a) displays the positions and the numbers of the laser run used in the analysis of the 100-200 pad-pitch device. For each of these runs, the measured and predicted percentages of the amplitudes are shown for the 4 pads in plots  (c)-(f): the black (red) points represent the calculated (measured) values, respectively. 

As an example, in run n.$6$, highlighted with an orange circle, the amplitudes in both pads 1 and 3 are $\sim 35\%$, as the laser is shot halfway between them, while pads 2 and 4 register smaller signals, $\sim 15\%$, since they are farther from the run position. In contrast, in run n.$10$, highlighted in green, pad 2 sees most of the amplitude as the laser is shot very close to it. 

Additionally, in Figure \ref{fig:laser_signalsharing} (b), the sum of the amplitude of the signals from the four read-out pads is plotted as a function of run number. This plot shows that the total signal amplitude is constant, regardless of the laser shot position.

Overall, the RSD LogA formula is a very powerful tool, allowing the prediction of signal sharing.

\begin{figure}[th!]
    \centering
    \includegraphics[scale=0.3]{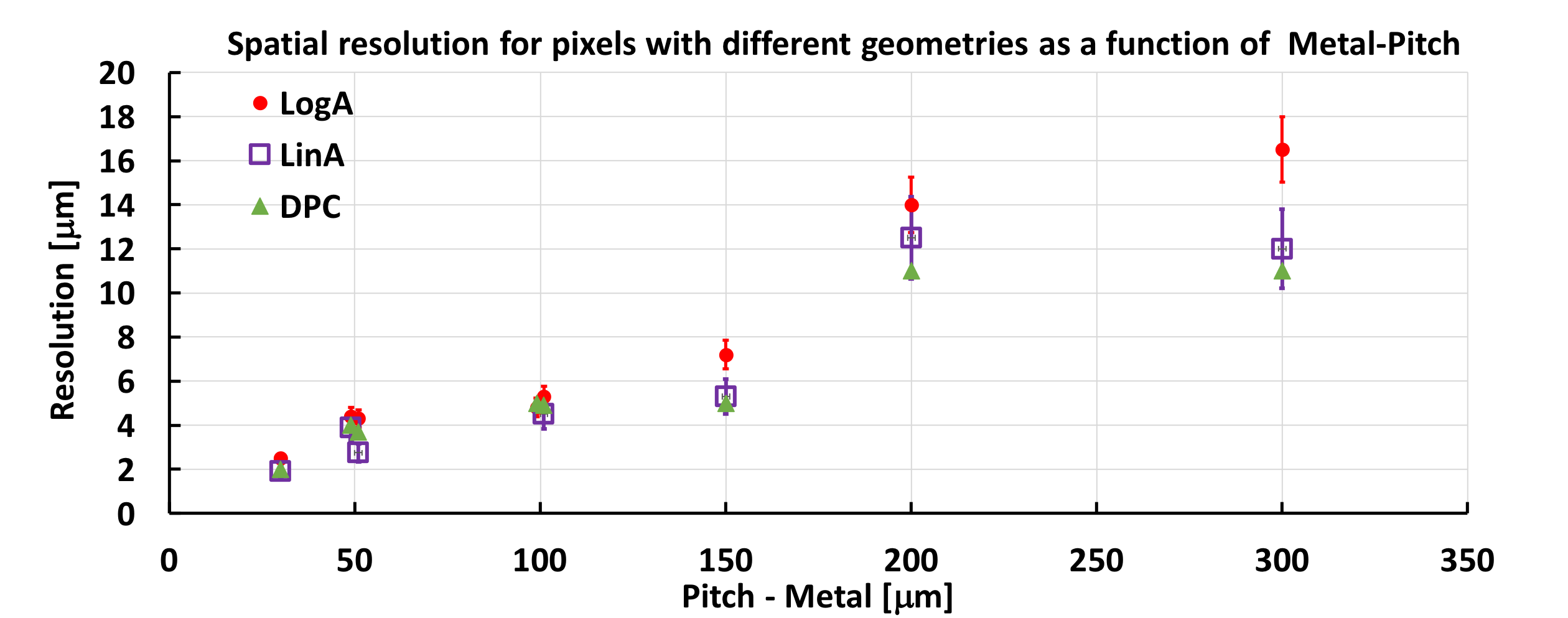}
    \caption{Spatial resolutions obtained with three different methods (LoqA, LinA, and DPC) for different RSD  geometries as a function of the interpad distance. }
    \label{fig:spaceres_laser}
\end{figure}

\subsection{Measurement of the spatial resolution}

unUsing the methods explained in section \ref{S:reco},  the spatial resolution has been computed for several geometries using the LogA, LinA, and DPC models. The results are reported in Figure \ref{fig:spaceres_laser} as a function of the pitch-metal distance (also called "interpad distance" in the following). 

In general, smaller pitch-metal distances yield a better spatial resolution. The resolution is extremely good, less than $5\;\mu m$ in both x and y direction, for interpad distances up to 100-150 $\mu m$. It is interesting that the LogA, LinA-tuned, and DPC-tuned models yield similar results up to interpads of about 200 $\mu m$, while above 200 $\mu m$ the LinA-tuned and DPC-tuned models become more performing. The poorer spatial resolution of the LogA model indicates that the assumption that the signal propagates in a triangular surface does not hold true over long distances. The LogA model predicts that far away pads see a larger fraction of the initial signal than what is actually measured (as visible in Figure \ref{fig:1Dampvsdist}.

\begin{figure}[th!]
    \centering
    \includegraphics[scale=0.16]{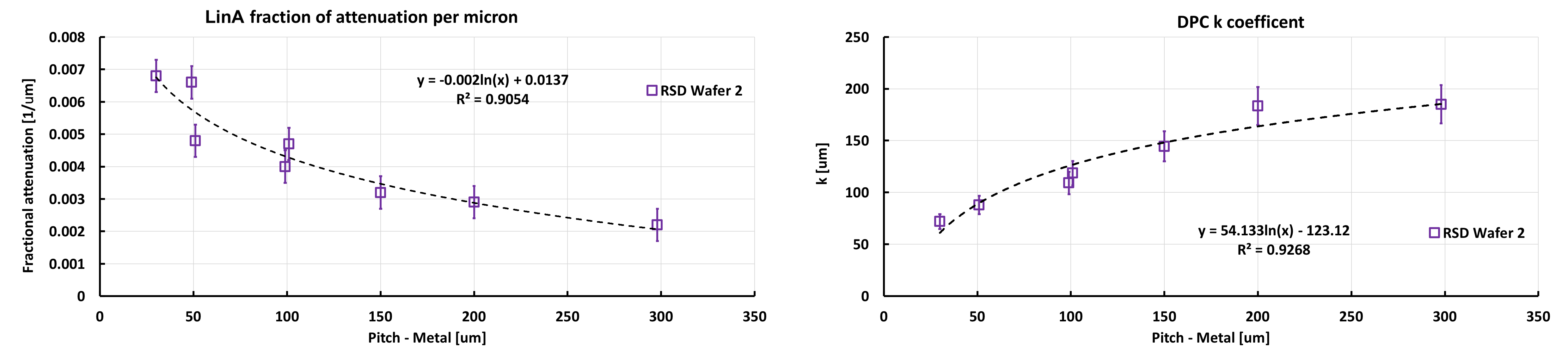}
    \caption{Left: LinA sharing law attenuation factor $\beta$ as a function of the pitch-metal distance. Right: DPC $k$ coefficient as a function of the pitch-metal distance. }
    \label{fig:RSD_lin_coeff}
\end{figure}

The results of the LinA and DPC models shown in Figure \ref{fig:spaceres_laser} have been obtained tuning the parameter $\beta$ (Equation (\ref{eq:LA})) and $k$ parameter, choosing the value that minimizes the spatial resolution for each geometry. The values of $\beta$ parameters used in this analysis are presented on the left side of Figure \ref{fig:RSD_lin_coeff} while the values of $k$ on the right side. 
The optimized $\beta$ and $k$ values have a logarithmic dependence on the interpad distance: this demonstrates that the attenuation is logarithmic with distance, as implemented in the LogA.

The same logarithmic dependence on interpad distance has been measured for the delay coefficient $\zeta$  of Eq.\ref{eq:lindelay}.

\begin{figure}[th!]
    \centering
    \includegraphics[scale=0.6]{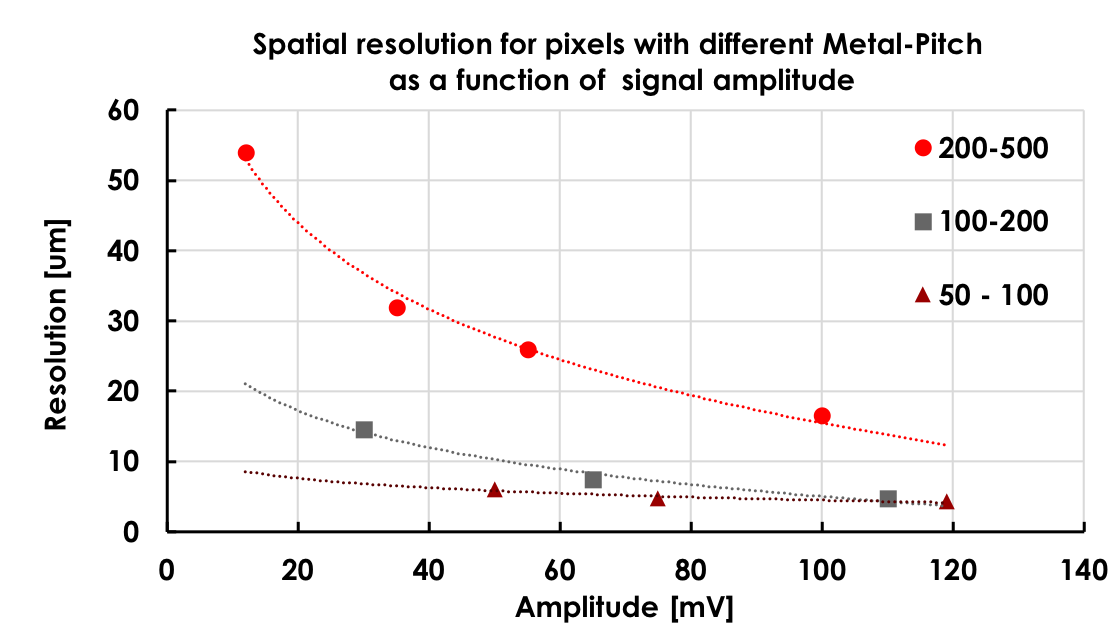}
    \caption{Spatial resolution as a function of the total signal amplitude for three pad-pitch geometries: 50-100, 100-200 and 200-500.}
    \label{fig:spaceresvsamp}
\end{figure}

The dependence of the spatial resolution upon the total signal amplitude is shown in Figure \ref{fig:spaceresvsamp}, for different geometries with the LogA model, as a function of the pitch-metal value. As expected, the resolution improves with the amplitude, most significantly in larger geometries. For small geometries, the resolution remains almost constant in a large range of amplitudes, showing that the performances are excellent even for low RSD gain. 

\subsubsection{Parameterization of the RSD spatial resolution.}

One interesting point is to understand why smaller geometries perform better. The position accuracy of an RSD sensor can be parameterized as the sum of three contributions: 

\begin{equation}
    \sigma^2_{tot}=\sigma^2_{jitter_x} + \sigma^2_{algorithm} + \sigma^2_{sensor}.
\end{equation}

\begin{itemize}
    \item $\sigma_{jitter_x}$: this term represents the spatial uncertainty induced by the electronic noise $\sigma_{el_{noise}}$  affecting the amplitudes.  Consider two simplified, with only 2 pads, RSDs with different interpad distances: 100 $\mu m$ and 300 $\mu m$. In these devices, a pad records the maximum signal amplitude (assumed to be 100 mV) when the signal is very near and a null amplitude when the signal is near the other pad. 
    The amplitude as a function of pad-hit distance for the two geometries is shown in Figure \ref{fig:jitter}.
    
    \begin{figure}[th!]
    \centering
    \includegraphics[scale=0.5]{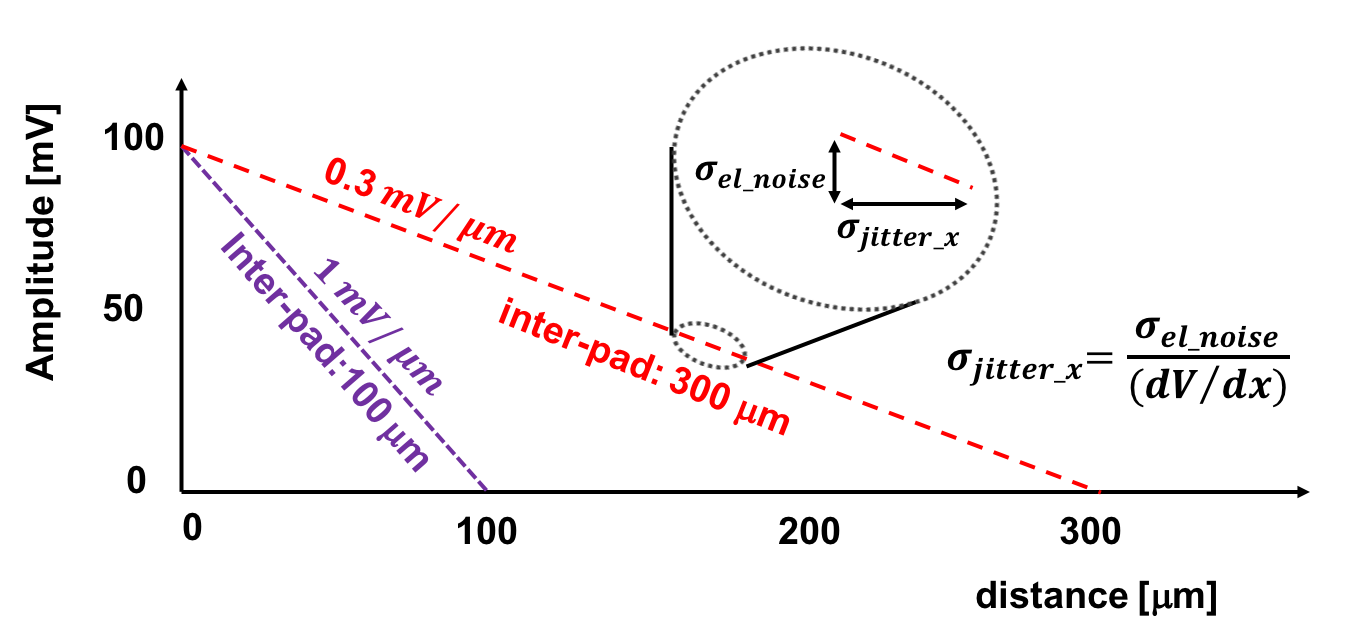}
    \caption{Amplitude as a function of distance for two RSD geometries. The jitter is worse for larger structures.}
    \label{fig:jitter}
\end{figure}
    
    In the 100 $\mu m$ geometry, the amplitude changes by $dV/dx=1 \;mV/ \mu m$, while in the 300 $\mu m$ geometry by $dV/dx=0.3 \; mV/ \mu m$. The electronic noise affecting the read-out of the amplitude will cause uncertainty in the localization of the hit given by
    
    \begin{equation}
   \sigma_{jitter_x} = \frac{\sigma_{el_{noise}}}{\frac{dV}{dx}}= \frac{\sigma_{el_{noise}}}{\frac{Amplitude}{Interpad}}
   \label{eq:jitter}
\end{equation}
  Assuming an electronic noise  of 3 $mV$, like in our setup, the jitter term on a 100 mV total signal is about  $\sigma_{jitter}=3 \; \mu m$   in the 100 $\mu m$ case, while it becomes  $\sigma_{jitter}=10 \; \mu m$ for the larger structure. The measured electronic noise was about 2 $mV$ for every metal pad geometry.

    \item $\sigma_{algorithm}$: the reconstruction code uses algorithms to infer the hit position from the measured signals. This can be done analytically, using a lookup table, or with more advanced techniques such as machine learning \cite{FS}. In all cases, the method selected has a given accuracy (whose impact has been mentioned in relation to Figure \ref{fig:spaceres_laser});
    
    \item $\sigma_{sensors}$: this term groups all sensor imperfections contributing to an uneven signal sharing among pads. The most obvious is a varying n+ resistivity: a 2$\%$ difference in n+ resistivity between two pads will unbalance the amplitude read out by the same amount, turning, for example, an equal 50 mV split, into 49 mV and 51 mV, respectively. Each reading will have an uncertainty $\sigma_{sensors} \sim 1\; \mu m$ for the 100 $\mu m$ geometry and $\sigma_{sensors} \sim 3\; \mu m$ for the 300 $\mu m$ design. 
\end{itemize}

In light of this parameterization, the values shown in Figure \ref{fig:spaceresvsamp} indicate that the spatial resolution of the largest structure (200-500) is dominated by the jitter term. For the two smaller geometries, the resolution becomes constant at high amplitude: in this region, the resolution is dominated by the uncertainties due to the sensor and/or reconstruction algorithm.


\subsection{Measurement of the timing resolution}

Once the hit position is reconstructed, the time of the hit is evaluated in a 2-step procedure. First, the time measured in each pad is corrected for the propagation delay from the hit position to the pad metal edge,  using Equation (\ref{eq:time}).


\begin{figure}[tbh]
    \centering
    \includegraphics[scale=0.31]{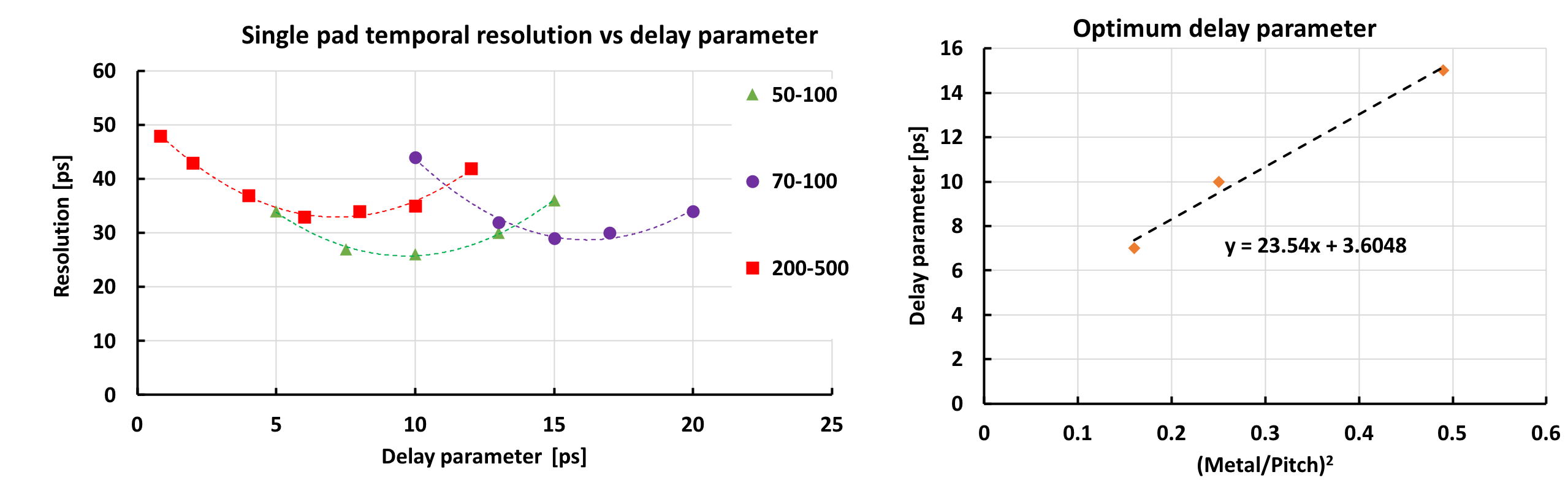}
    \caption{Left side: Single pad temporal resolution as a function of the delay parameter for three pad-pitch geometries: 50-100, 70-100 and 200-500. Right side: Optimum delay parameter as a function of the ratio $(metal/pitch)^2$.}
    \label{fig:delay}
\end{figure}

Figure \ref{fig:delay}, plot on the left, shows how the temporal resolution of a single pad changes as a function of the delay parameter $\gamma$, for three different geometries from the same wafer; the plot on the right  shows how the optimized delay parameter depends linearly on the ratio $(metal/pitch)^2$. This fact is an indication that the geometrical properties of a given design, and not just the $n^+$ resistivity, play a role in the delay.

\begin{figure}[tbp]
    \centering
    \includegraphics[scale=0.3]{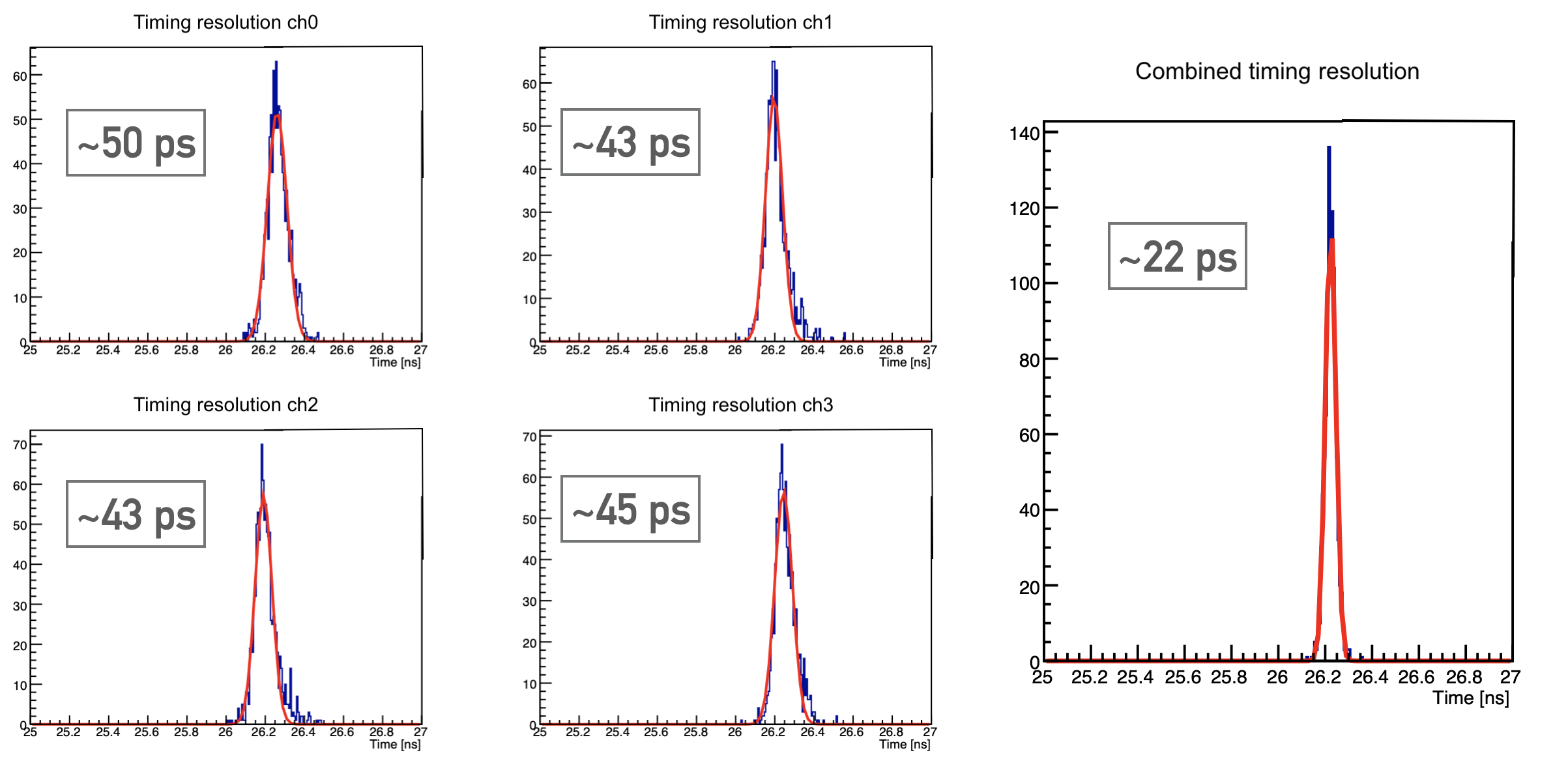}
    \caption{On the left: single-channel temporal resolution for the four read-out pads for laser data. On the right: total temporal resolution obtained combining the four pads timestamps.}
    \label{fig:timeres_100-200_laser}
\end{figure}

In a second step, the time of the event is defined as the amplitude-weighted time average of the active read-out pads. The temporal resolution is obtained as the sigma of the $t_{trigger}-t_{reco}$ distribution. Figure \ref{fig:timeres_100-200_laser}  shows the temporal resolution of each of the 4 active pads for the 100-200 geometry (on the left) and the obtained combined resolution (on the right). The resolution of the single channel is about $\sim 45$ ps, while the combination of the 4 channels yields to about $\sim 22 \; ps$, as expected from measurements with uncorrelated noise ($22\; ps \sim 45\;ps/\sqrt{4}$).

\begin{figure}[tbh]
    \centering
    \includegraphics[scale=0.32]{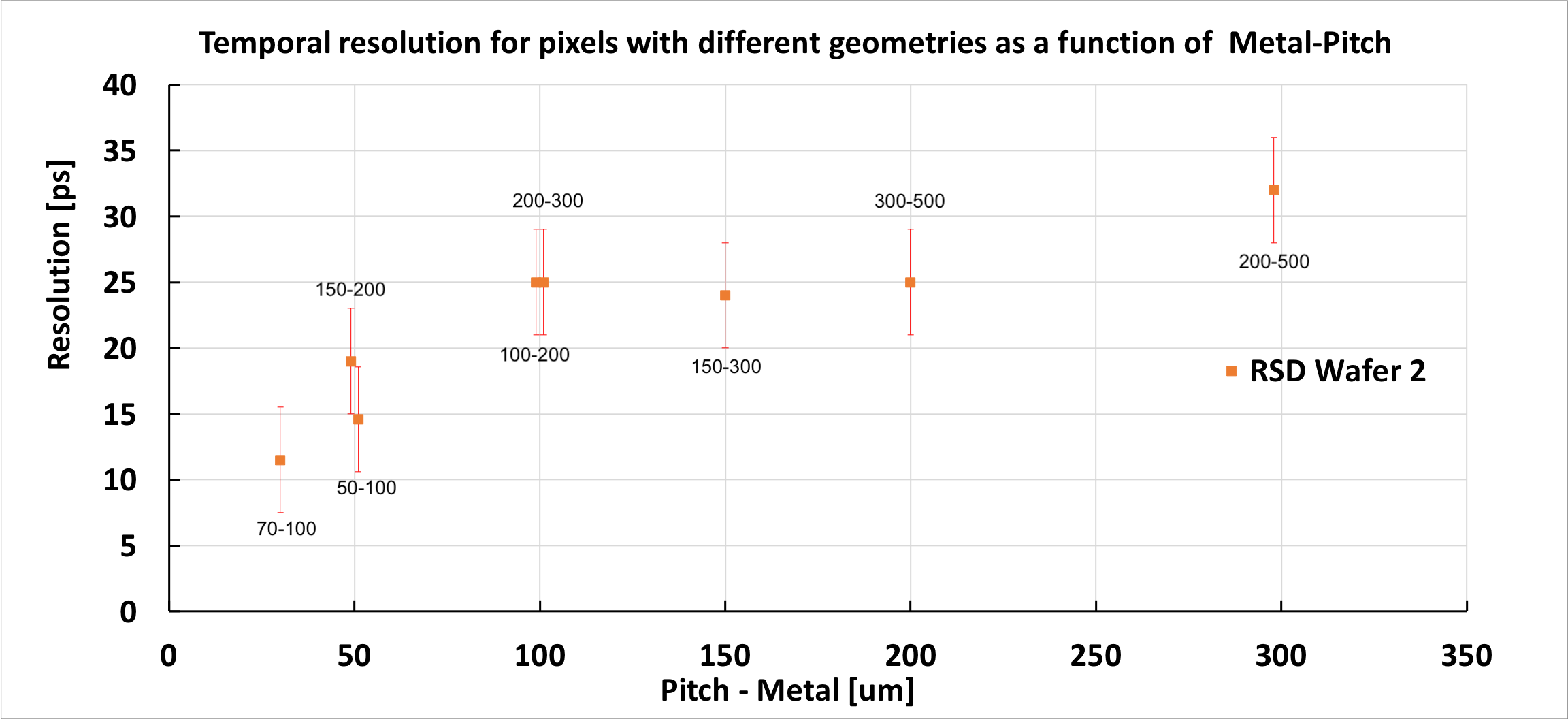}
    \caption{Temporal resolution for different RSD pad-pitch geometries as function of the interpad distance.}
    \label{fig:timeres_laser}
\end{figure}

The temporal resolution as a function of the interpad distance is shown in Figure \ref{fig:timeres_laser}. The resolution increases up to $pitch-pad=100\;\mu m$, reaching a constant level around $25-30\; ps$ for larger geometries. 

\subsection{Summary of measurements with laser TCT}

Table \ref{tab:spaceres} summarizes the results obtained with the laser measurements for each geometry. 
The temporal resolution is in line with the best LGAD results obtained with laser signals (no Landau fluctuations), demonstrating how the RSD design does not degrade the LGAD performances. The RSD spatial resolution is extremely good for all geometries as the reconstruction method is able to exploit the power of internal signal sharing.  
 
\begin{table}[h!]
    \centering
    \begin{tabular}{c|c|c}
         pad-pitch geometry & spatial resolution [$\mu m$] & temporal resolution [ps] \\
        \hline
        50-100  & 4.3 & 14.7\\
        70-100 & 2.5 & 11.5\\
        100-200 & 4.8 & 25\\
        150-200 & 4.4 & 19\\
        150-300 & 7.2 & 24\\
        200-300 & 5.3 & 25\\
        200-500 & 16.5 & 32\\
        300-500 & 14 & 25\\
    \end{tabular}
    \caption{Spatial and temporal resolutions from TCT measurements for different RSD pad-pitch geometries. These results refer to studies where the laser has been shot in the interpad region of the sensors, as shown in Figure \ref{fig:laser} }
    \label{tab:spaceres}
\end{table}

\section{RSD simulation in Weightfield2}

\begin{figure}[th!]
    \centering
    \includegraphics[scale=0.5]{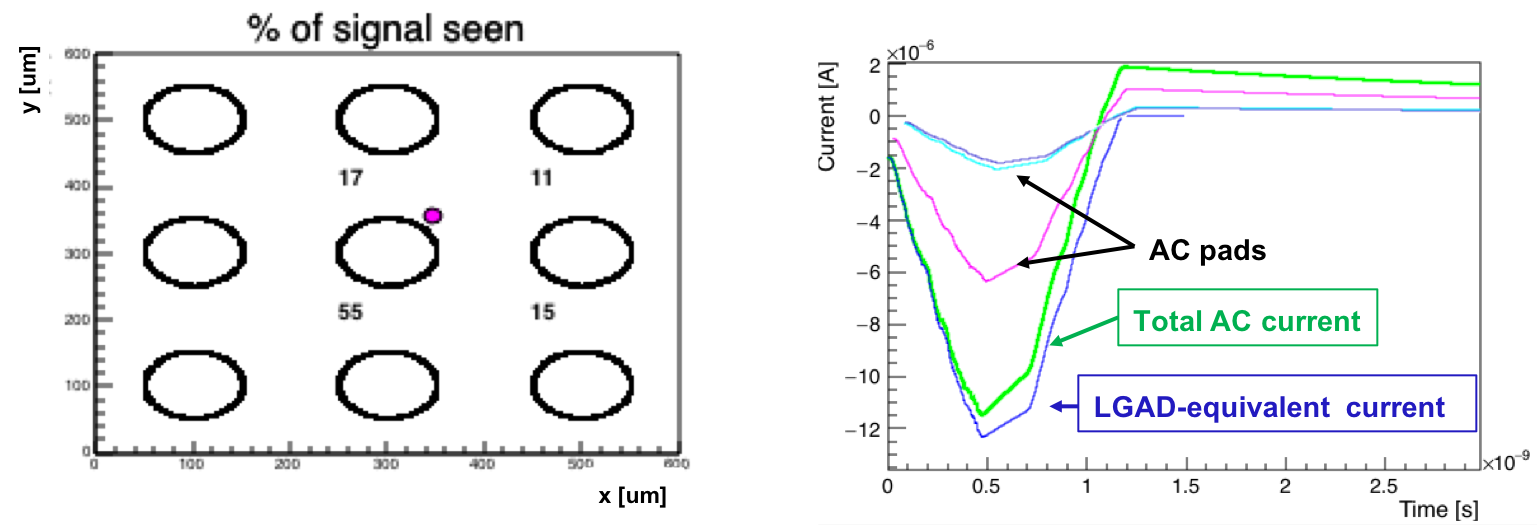}
    \caption{An example of pad geometry and of the current signals simulated by WF2-RSD.}
    \label{fig:wf2}
\end{figure}

Weightfield2\footnote{Shareware at http://cern.ch/nicolo} is a simulation program that has been extensively used in the design and characterization of the LGAD properties \cite{WF2}. WF2 emulates the passage of a particle in a silicon detector and generates the output current, including the effect of non-uniform ionization, gain, geometry, and acceptor removal. The RSD principle of operation has been added to the WF2 program by implementing the prediction of the RSD LogA formulas, Equations (\ref{eq:masterformula}) and (\ref{eq:time}). In WF2-RSD it is possible to select a given pad geometry and to simulate the current signals in the nearby pads as a function of the hit position.  Figure \ref{fig:wf2} shows on the left an example of geometry and indicated by the purple circle the position of the hit. The number near each pad is the fraction of the signal seen in that pad. On the right side, the shape of the DC current, of the total AC current, and the currents in each of the pad is shown. The program allows generating many events in batch mode and writing the events to a file for offline analysis. 

\section{Beam test measurements}
Two $3\times3$ RSD matrices from the same wafer, with $100-200$ and $190-200$ pad-pitch geometry, operating at -430 V, were measured at the Fermilab Test Beam Facility. In this facility, 120 GeV/c protons are delivered every minute in 4 seconds spills, each containing 50-100k particles. RSDs were mounted to a block cooled by circulating glycol that enables holding the sensor at a constant temperature of $ (22 \pm 0.1) ^\circ C$. They were wire-bonded to a $16$-channel read-out board, designed at Fermilab to test sensors with sizes as large as $8.5\times8.5 \rm{mm}^2$ at voltages up to $1\;kV$. 
Each read-out channel consists of a 2-stage amplifier chain based on the Mini-Circuits GALI-66+ integrated circuit. In this particular configuration,  amplifiers used a $25 \; \Omega$ input resistance, a $\sim 5 \; k\Omega$ total trans-resistance, and a bandwidth of $1 \; GHz$.
The box containing the device under test (DUT) was preceded by a tracking telescope, approximately 2 meters upstream along the beamline. The tracking system was provided with 14 layers of strips and 4 layers of pixels, followed by two additional layers of strips, offering a spatial resolution of $\sim 45\; \mu m$, somewhat degraded by the extrapolation downstream and by material between the telescope and the DUT. Additionally, the setup was instrumented with a Photek 240 micro-channel plate detector, placed inside the environmental chamber after the DUTs, with a temporal resolution better than 10 ps \cite{FNAL}. 
 
The data were acquired with a Keysight MSOX92004A 4-channels digital scope with 40 GS/s and 20 GHz analog bandwidth. Given that only 4 signals could be recorded simultaneously, different types of datasets were stored:
\begin{itemize}
    \item 4 AC pads,
    \item 3 AC pads and the Photek \cite{MT2},
    \item 2 AC pads, the DC pad, and the Photek.
\end{itemize}

The following quality cuts were required in the analysis: 
\begin{itemize}
    \item the presence of only one isolated proton per event in the tracking system; its track must have hits at least in 14 layers of the tracking telescope and in one of the two downstream strip layers. 
\item the signal on each RSD read-out pad has to be larger than 15 mV, 
\item  the signals should not saturate the scope 
\item the sum of the AC signals has to be larger than 80 mV (60 mV) when read with 4 (3) pads.
\end{itemize}

\label{S:3}
\subsection{$100-200$ geometry: 4-pads configuration studies}
Since the tracker resolution, approximately $45 \; \mu m$, is large compared to the RSD resolution, it is not possible to perform a simple comparison of the tracker and RSD positions event by event. 

\begin{figure}[th!]
    \centering
    \includegraphics[scale=0.14]{ 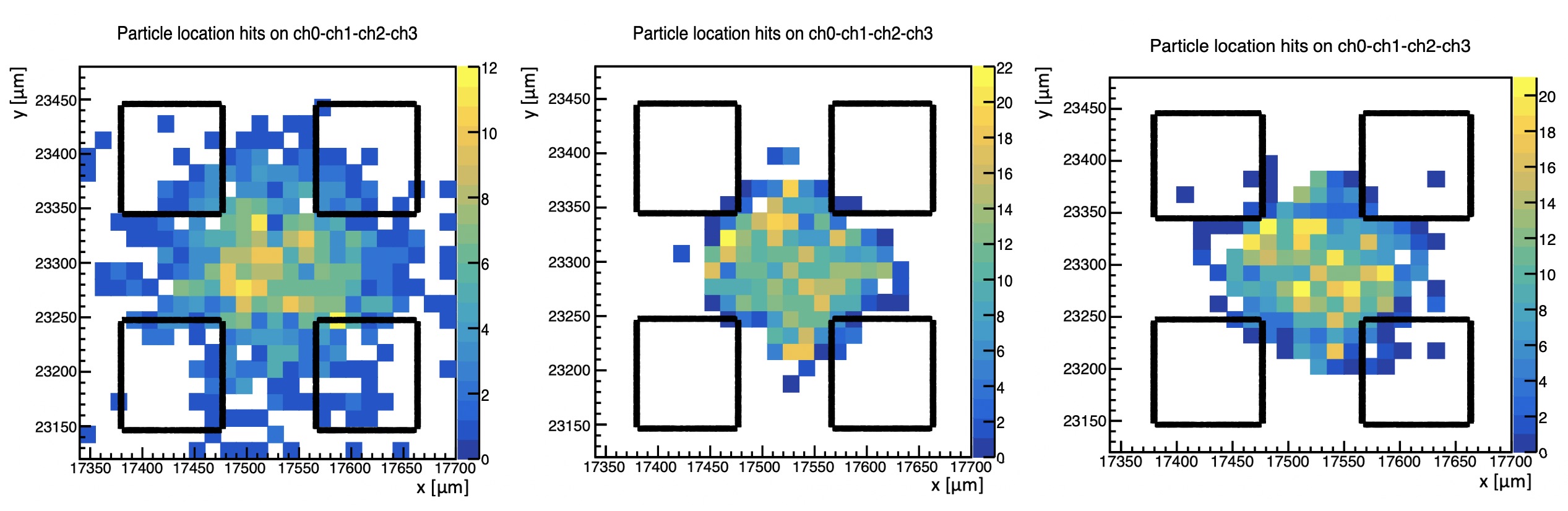}
    \caption{Hit maps for the 100-200 geometry represented using the tracker coordinates (left) and reconstructed with the analytical method using the LogA (center) and LinA (right) models.}
    \label{fig:hitmap_100-200_RSD}
\end{figure}

Using the selection cuts above, Figure \ref{fig:hitmap_100-200_RSD} reports the hit position map using the tracker (left), the LogA (center), and LinA (right) models. The area covered by the hits is larger when using the tracker, having a worse resolution, while LogA and LinA predict similar distributions. 
 
\begin{figure}[th!]
    \centering
    \includegraphics[scale=0.13]{ 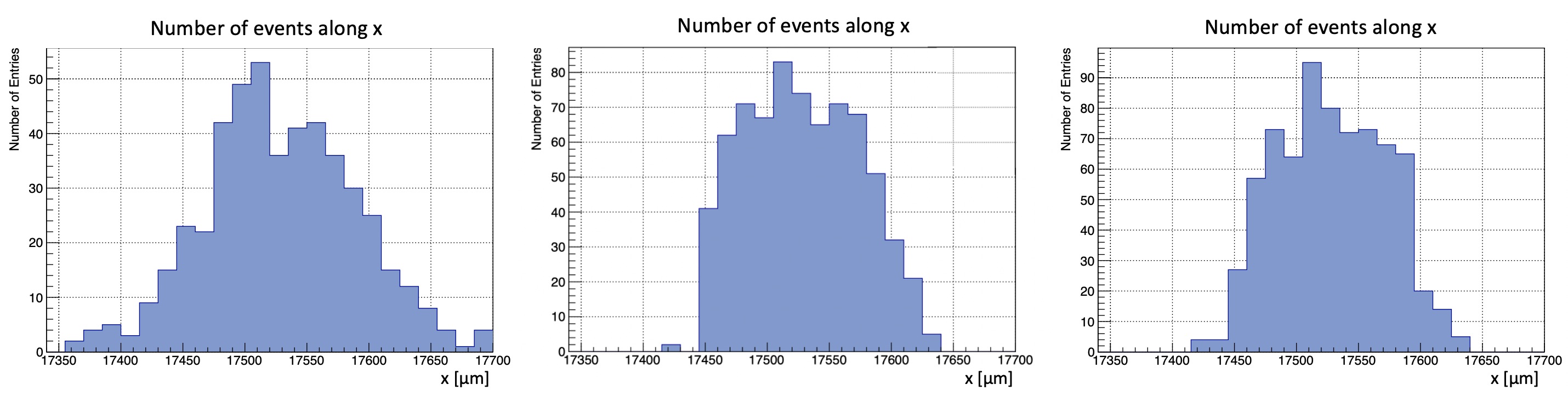}
    \caption{Number of events represented as a function of x-coordinate in the range  $23255\; \mu m < y < 23345\; \mu m$ for the tracker (left), the LogA (center) and LinA (right) models.}
    \label{fig:projx}
\end{figure}

Given that the beam illuminates the DUT uniformly, the density of hits as a function of position is constant. However, the measured distribution includes the reconstruction smearing.  This effect is shown in Figure \ref{fig:projx}, where the events in the y-range $23255\; \mu m < y < 23345\; \mu m$ are plotted versus the x-axis. For the tracker case, the distribution is not flat, as expected considering its resolution, while both the LogA and LinA methods show a good spatial uniformity, showing that the smearing introduced by the resolution of the RSD device is smaller. 

\begin{figure}[th!]
    \centering
    \includegraphics[scale=0.14]{ 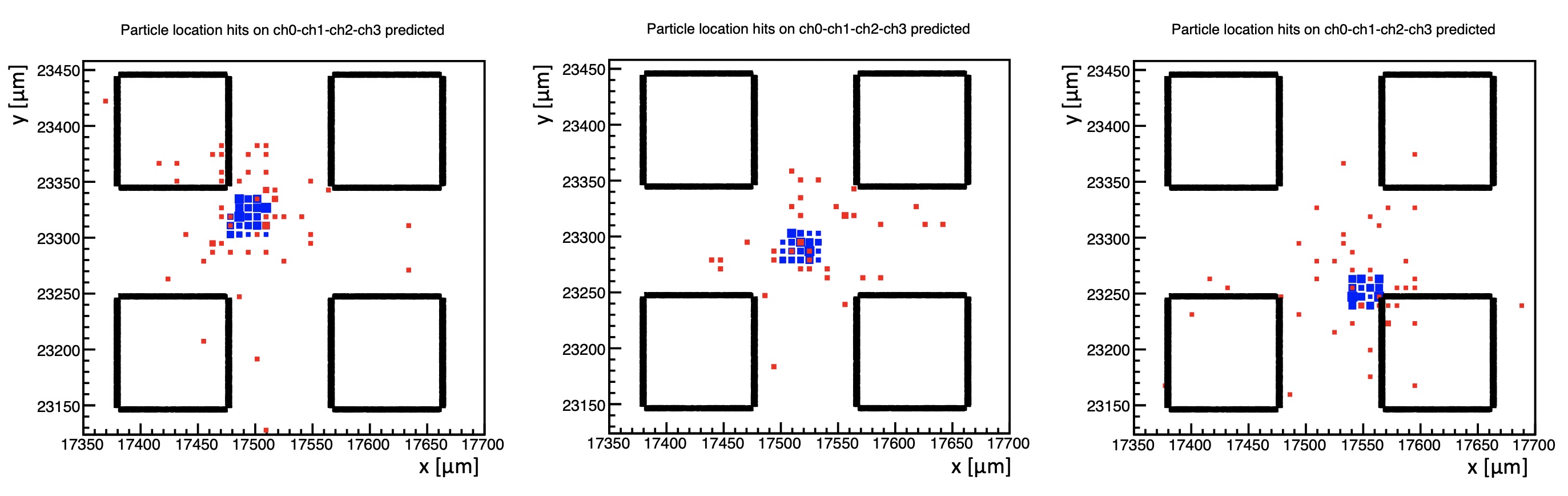}
    \caption{Hit maps comparing hit positions defined by RSD coordinates (blue) and the tracker ones (red) for events selected in $30\times30\;\mu m^2$ defined by the RSD coordinates in the interior of the 4 pads.}
    \label{fig:squareRSDvstracker}
\end{figure}

Even though an event-by-event comparison between the tracker and the RSD position reconstruction cannot be made, a comparison between mean values is possible. Figure \ref{fig:squareRSDvstracker} shows where the tracker reconstructed positions are (red points) when selecting the events in three specific areas (blue points) using the RSD position. 

\begin{figure}[th!]
    \centering
    \includegraphics[scale=0.17]{ 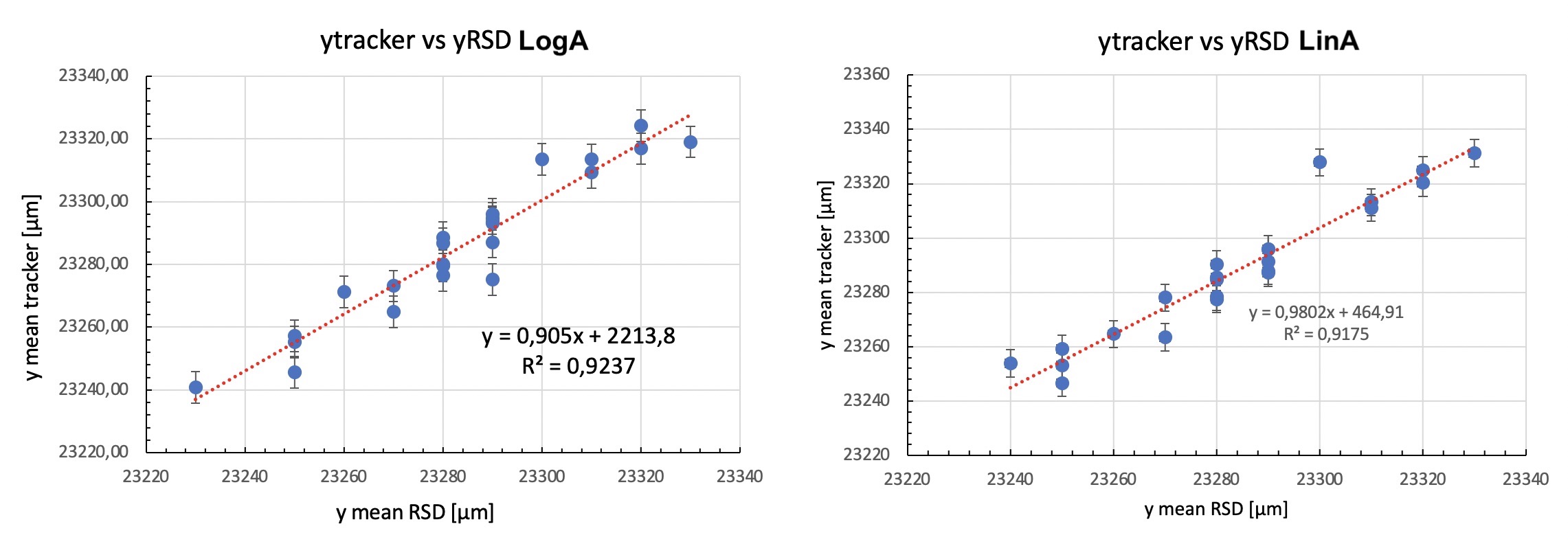}
    \caption{$y_{tracker}^{mean}$ plotted as function of $y_{RSD}^{mean}$ for the LogA (left) and the LinA model. The parameters resulting from the linear fit are displayed on the plots.}
    \label{fig:yRSDvstracker}
\end{figure}

The systematic comparison between mean values is shown in  Figure \ref{fig:yRSDvstracker}, where $y_{tracker}^{mean}$ is plotted as a function of $y_{RSD}^{mean}$.  For each point, the events are selected by requiring their RSD position to be within a square of $30\times30 \; \mu m^2$. Both LogA (left) and LinA (right) models work very well; the correlation factors $R^2$ and the fit slope are close to 1. 

These comparisons prove that the LogA model can accurately reproduce experimental data despite the absence of tunable parameters. In the LinA model, it is instead necessary to optimize the attenuation factor $\beta$, whose best value in these studies is $\beta=0.003/\mu m$.

\subsubsection{Spatial resolution using signals from 3 or 4 pads}

In the dataset used for timing measurements, only 3 AC pads have been readout. The analysis presented in the previous section is thus repeated, including only 3 pads to compare the 4- and 3-pads spatial resolutions. If we consider events in a triangular area near the included pads, the spatial resolution for the 3-pads dataset is equal to the 4-pads one. Moving to the total device area, the spatial resolution worsens due to the missing pad.

\subsection{$100-200$ geometry: timing studies using  3 AC pads and the Photek detector as time reference}

The first step in the timing analysis is to measure the propagation delay, from the hit point to the read-out pads.  The hit positions are reconstructed using the LogA model.

\begin{figure}[th!]
    \centering
    \includegraphics[scale=0.16]{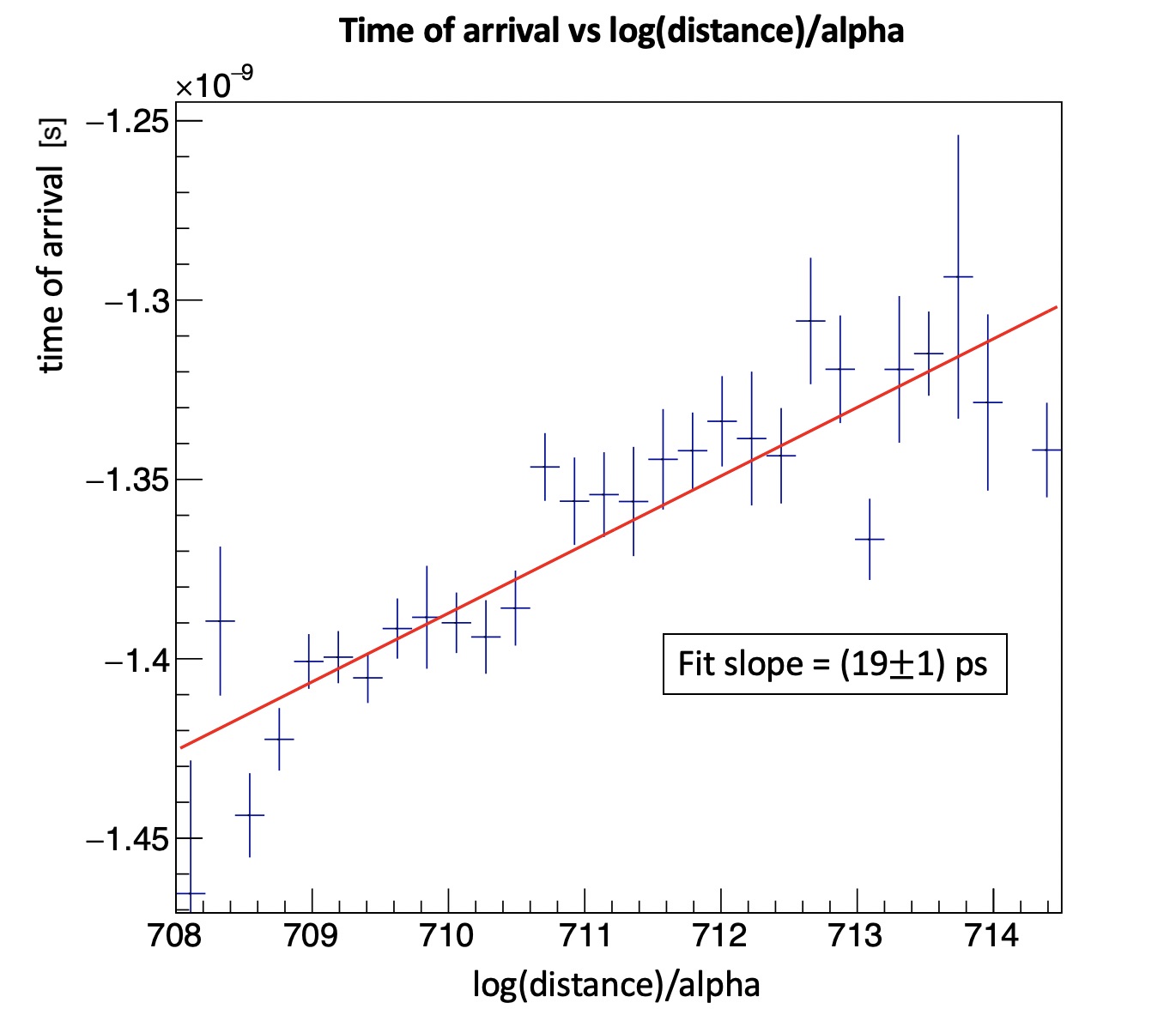}
    \caption{Time of arrival of the 100-200 3-pad configuration as a function of $\frac{ln(d/{d_0})}{\alpha}$.}
    \label{fig:CFD_RSDvstracker}
\end{figure}

\begin{figure}[th!]
    \centering
    \includegraphics[scale=0.2]{ 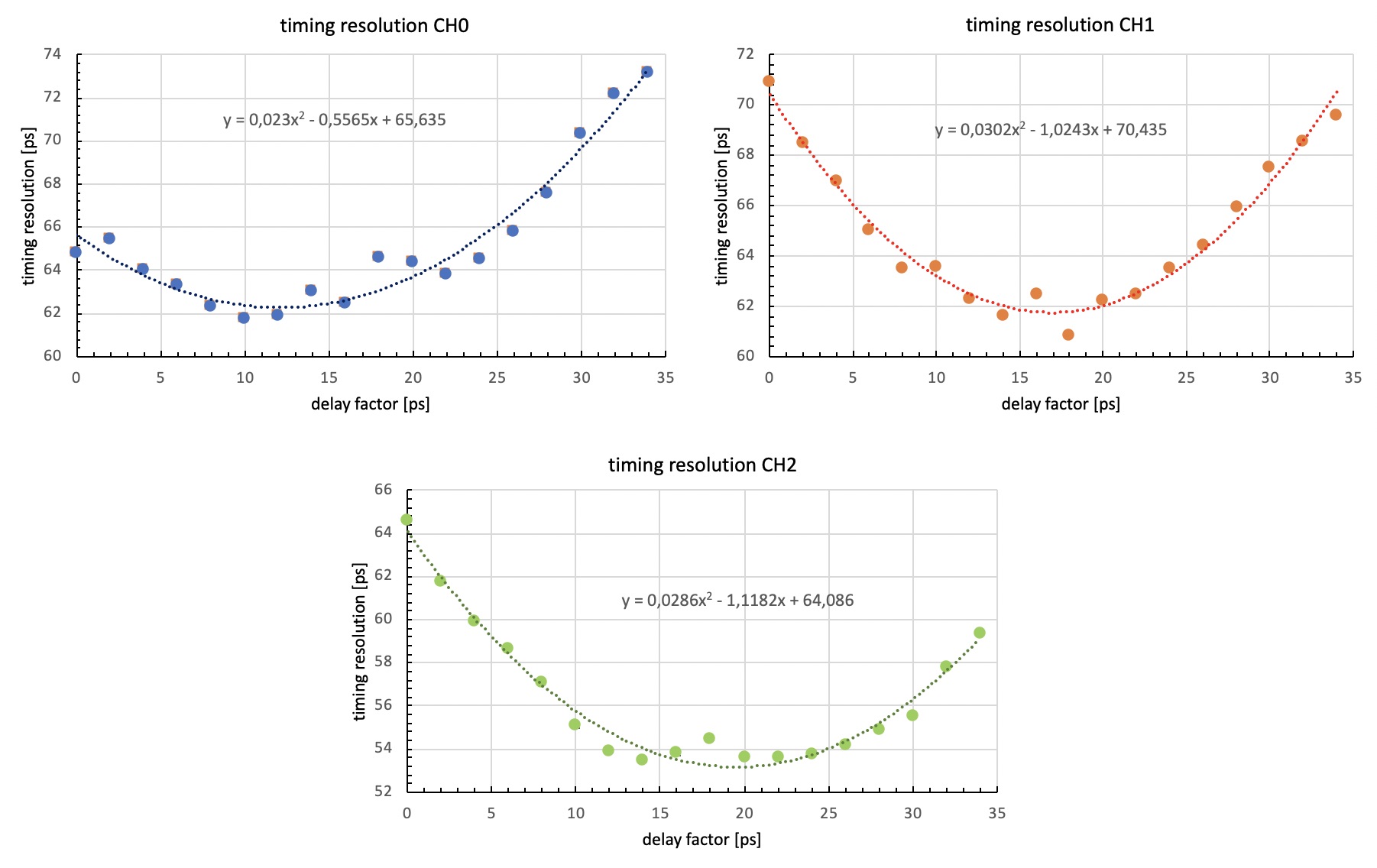}
    \caption{Time resolution represented as a function of time delay parameter $\gamma$ for the three RSD channel of the 100-200 3-pads configuration.}
    \label{fig:timeres_singlechannel}
\end{figure}

Two methods are used to extract the best time delay parameter $\gamma$ (Equation (\ref{eq:time})):
\begin{itemize}
\item Direct method: the time of arrival is plotted as function of $\frac{ln(d/{d_0})}{\alpha}$ and fitted linearly. The fit slope represents the time delay parameter, here $\gamma_{dir}=(19 \pm 1)\;ps$ (Figure \ref{fig:CFD_RSDvstracker}). 

\item Indirect method: the time delay $\gamma$ is found by minimizing the temporal resolution as a function of the delay parameter itself,  as shown in Figure \ref{fig:timeres_singlechannel}. For each pad, the  best values are $\gamma_{CH0}=(12.1 \pm 0.6)\;ps$, $\gamma_{CH1}=(17.0 \pm 0.1)\;ps$, and $\gamma_{CH2}=(19.5 \pm 0.2)\;ps$. 
\end{itemize}

The time delay value used in the analysis is the mean of the four results from the two methods, while as uncertainty the (max-min)/2 is used: $\gamma=(16.9 \pm 3.7)\;ps$. 

\begin{figure}[th!]
    \centering
    \includegraphics[scale=0.2]{ 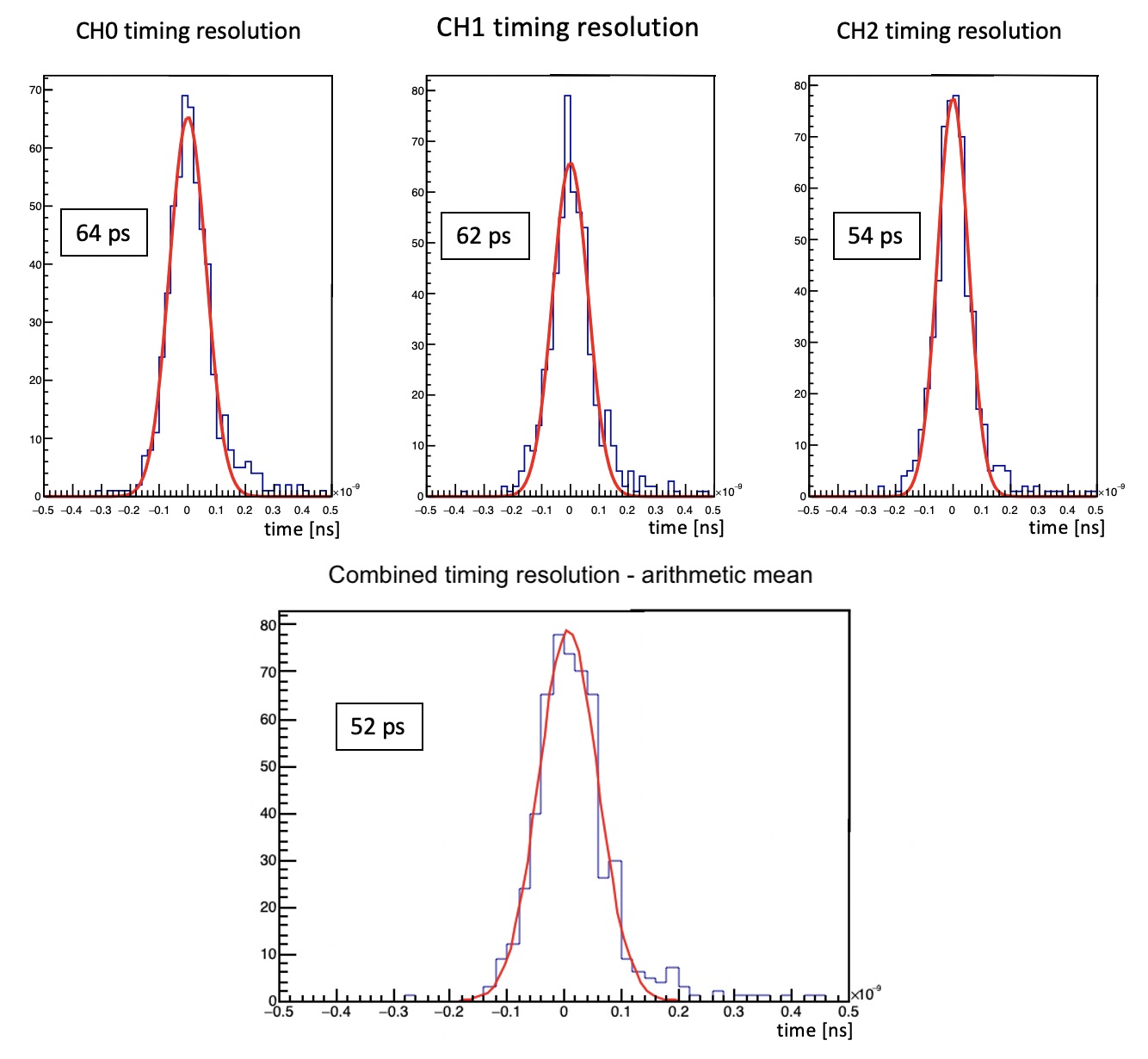}
    \caption{Top: single channel temporal resolution for the three read-out pads at test beam. Bottom: total temporal resolution obtained as the arithmetic mean of the three pads timestamps.}
    \label{fig:timeres_100-200_testbeam}
\end{figure}

Once the time delay parameter has been evaluated, the time information from the 3 pads can be combined in a  arithmetic mean to form the hit time of arrival $t^{a}_{RSD}$.  The temporal resolution is measured as the sigma of the distribution of $t^{a}_{RSD}-t_{Photek}$. 
Figure \ref{fig:timeres_100-200_testbeam} shows the three single-channel resolutions (top) and the combined one (bottom),  $\sigma_{tot} =(51.9 \pm 0.5)\;ps$. In contrast to the laser results (Figure \ref{fig:timeres_100-200_laser}), this value is higher than the one expected from combining three single-channel resolutions with uncorrelated uncertainties ($\sim 34\; ps$). 
In a beam test, the temporal resolution is mainly determined by the jitter term, deriving from the electronic noise of each read-out channels, and by the non-uniform ionization term\cite{ROPP}. In RSDs, the signals spread from the hit point to the AC pads are a "copy" of the same signal, and for this reason, their shape is largely correlated. As a consequence, also the non-uniform ionization term to the resolution is correlated among pads, and the combination of multiple pads does not lead to an improvement of this contribution.

\begin{figure}[bph!]
    \centering
    \includegraphics[scale=0.18]{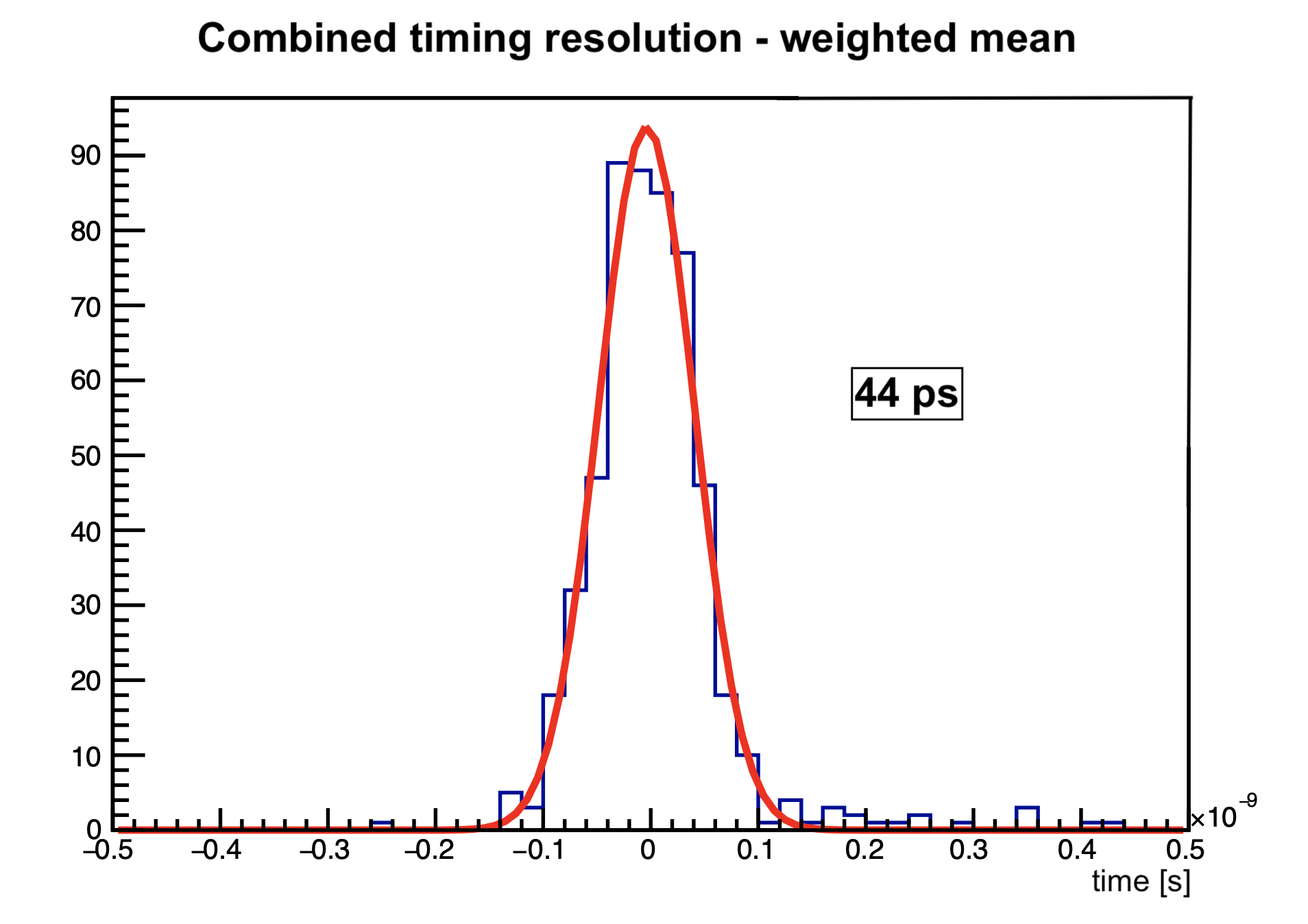}
    \caption{$t_{RSD}-t_{Photek}$ distribution with $t_{RSD}$ calculated as the amplitude-weighted mean of the three pads timestamps.}
    \label{fig:weighted_timeres}
\end{figure}

The combined timestamp $t_{RSD}$ used in  
 Figure \ref{fig:timeres_100-200_testbeam} (bottom plot) has been calculated as the arithmetic average of the three single-channel RSD timestamps. The temporal resolution  can be improved if $t_{RSD}$ is instead calculated as an amplitude-weighted average, $t^w_{RSD}$, yielding to a resolution of $\sigma_{tot} = (44.4 \pm 0.3)\;ps$ (Figure \ref{fig:weighted_timeres}). This value is comparable with the LGAD performances obtained at the same gain and using the same electronics. The effect of amplitude weighting is to reduce the importance of small signals, where the signal-to-noise ratio is worse.

\subsection{190-200 geometry: studies using 3 AC pads and the Photek}

The RSD with 190-200 geometry was used to study the effect of large metal pads and floating electrodes on the signal reconstruction. 

\subsubsection{Position reconstruction and the effect of floating pads}

\begin{figure}[thbp]
    \centering
    \includegraphics[scale=0.13]{ 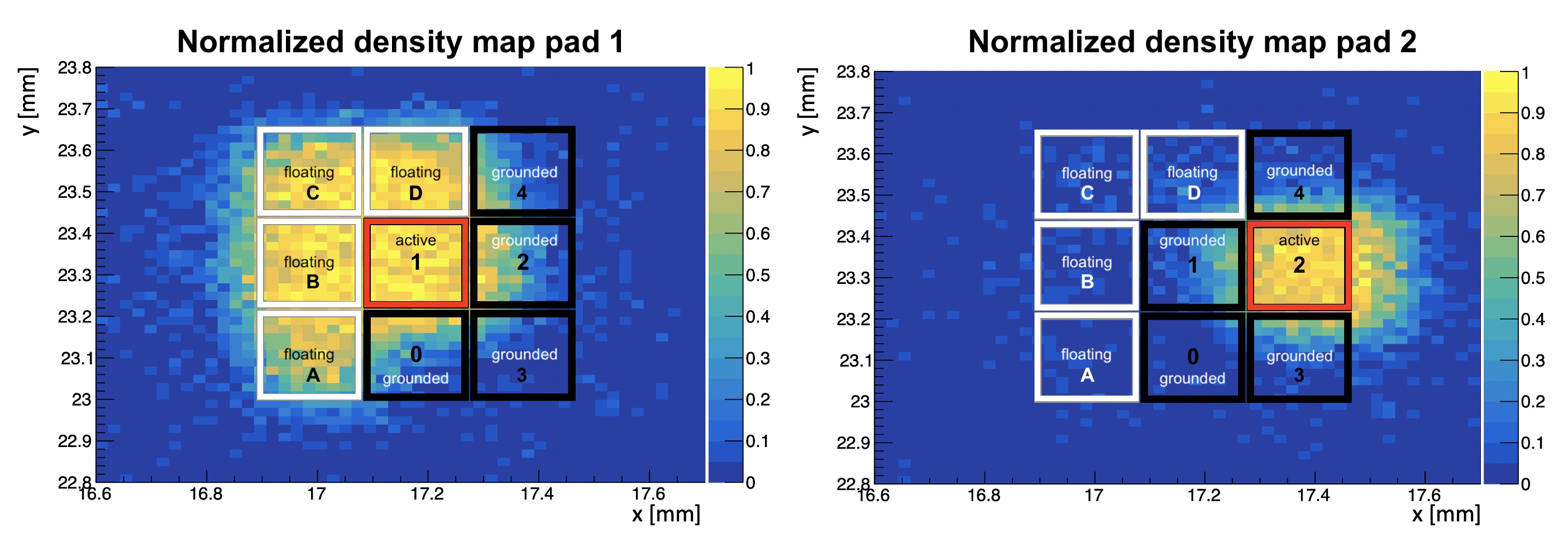}
    \caption{Normalized density maps for pad 1 (left) and pad 2 (right) of the 190-200 geometry. The red square is the read-out pad, black squares indicate grounded pads, while the white ones represent the floating pads.}
    \label{fig:efficiency_190-200}
\end{figure}

Figure \ref{fig:efficiency_190-200} shows the normalized density map of hits reconstructed by the tracker when the RSD pad 1 (left side) or pad 2 (right side) is active. Pad 1 has 4 floating neighbours, pads A, B, C, and D  and four grounded neighbours, pads 0, 2, 3, and 4.
Clearly, when the particles hit a floating pad, the signal is well visible in pad 1. In contrast, when particles hit a grounded pad, there is no signal sharing.  The analysis of the map for pad 2 (plot of the right side) brings to similar conclusions: almost all neighbours of pad 2 are grounded, and hits are visible only when they directly occur on the pad.  


\begin{figure}[th!]
    \centering
    \includegraphics[scale=0.14]{ 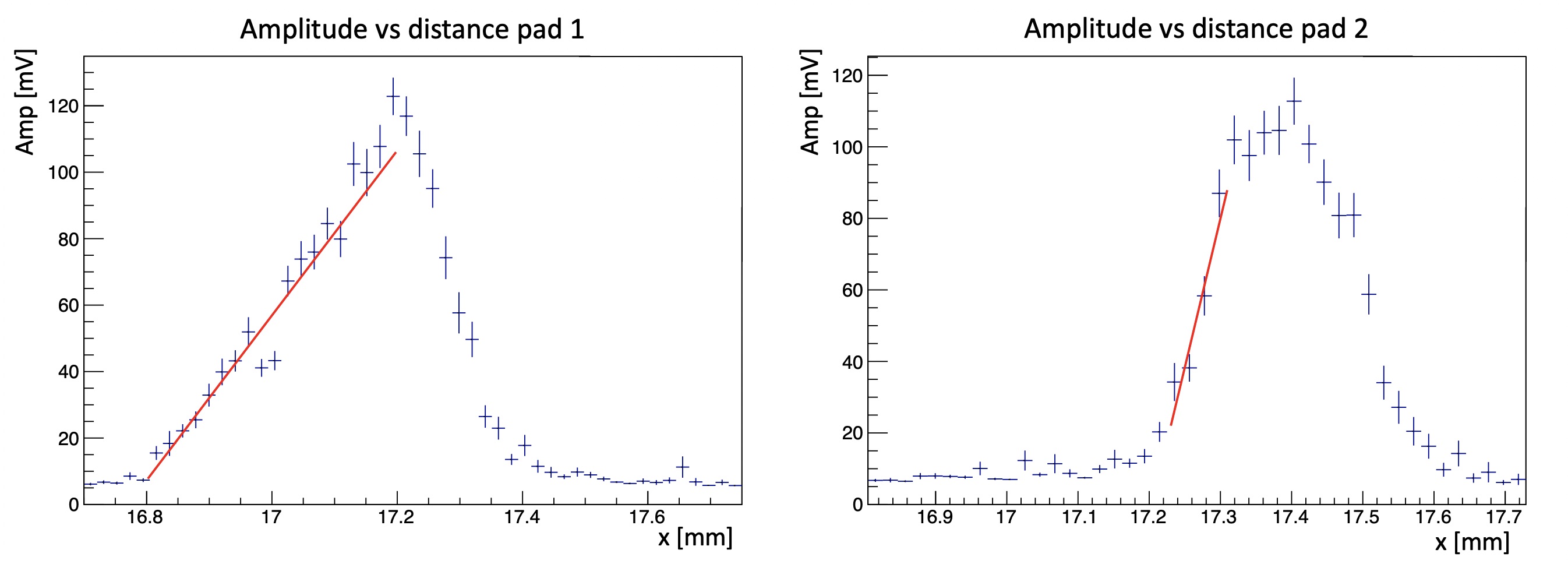}
    \caption{Amplitude along x axis for pad 1 (left) and 2 (right): the first is amidst a grounded and a floating pad, while the second is between a grounded pad and the sensor edge.}
    \label{fig:ampvsx190-200}
\end{figure}

For the same two AC pads, Figure \ref{fig:ampvsx190-200} shows the amplitude of the signal as a function of the x-coordinate. Pad 1 (left) is between a grounded pad and a floating one, while pad 2 (right) is placed between a grounded pad and the device edge. The attenuation of the signal is steeper next to grounded pads than underneath the floating ones. 

These plots highlight how RSD with large metal pads works differently with respect to RSD with small metal pads: 
\begin{itemize}
    \item most of the events are seen by only one pad,
    \item signal sharing is limited in a very narrow region between pads, 
    \item they behave similarly to traditional LGADs but benefit from a design that provides a  100\% fill factor.
\end{itemize}

The large majority of events recorded in this dataset is thus seen only by one of the read-out pads. Hit reconstruction is not applied to the 190-200 matrix, and the spatial resolution is estimated as $pad\:size/\sqrt{12}$ ($\sim 55 \;\mu m$). 

\subsubsection{Temporal resolution}
As a consequence of the very limited signal sharing, the temporal resolution of this geometry is very good for 1-pad events (very small jitter since the signal is large) and slightly worse for the 2-pads events.  
For this specific analysis, an extra cut on the tracker hit coordinates is applied, requiring the events to have $x>17.05\;mm$ and $y<23.45\;mm$.

The temporal resolution obtained is  $(\sigma = 32\pm 1)\;ps$ for the 1-pad events, and $(\sigma = 42.1\pm 0.6)\;ps$ for the 1-, 2- and 3-pads events. 

Overall, RSD sensors with the AC metal pads of about the same size of the pitch offer an excellent temporal resolution and a 100$\%$ fill factor. They represent a good choice when the desired spatial resolution is of the order of (pixel size)$/\sqrt(12)$.

\subsection{100-200 geometry: studies of the DC pad signals}

In this section, the properties of the RSD signal generated on the $n^+$ layer are analyzed by concurrently reading the currents from an AC pad and the $n^+$ layer. The $n^+$ layer is readout by contacting the metal connected to the DC contact, as shown in \ref{fig:RSD}. The DC contact extends around the periphery of the of the $n^+$ layer. 
This study has identified two families of events, depending on the impact point of the particle on the RSD (Figure \ref{fig:DCpulse}):

\begin{figure}[th!]
    \centering
    \includegraphics[scale=0.22]{ 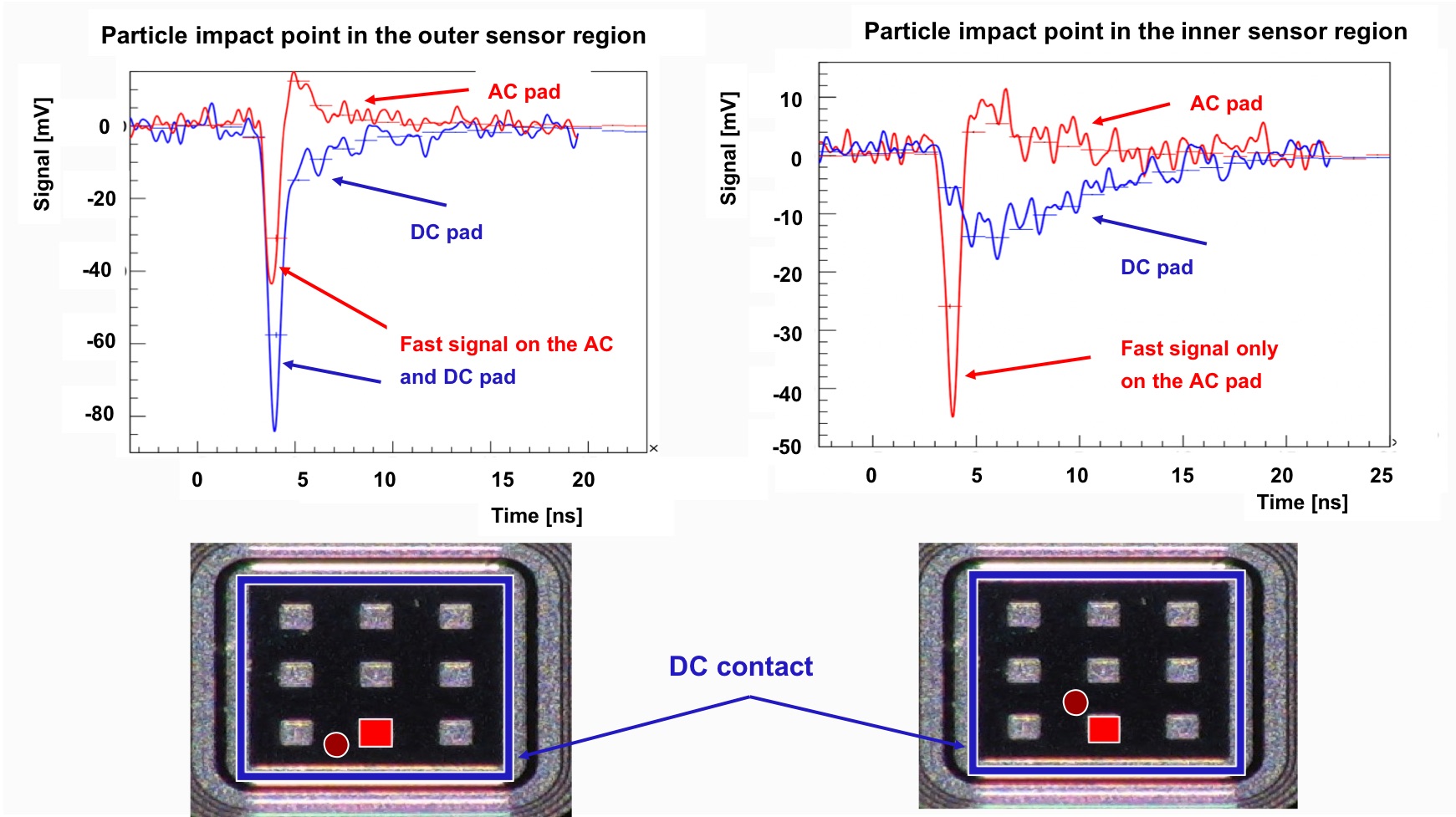}
    \caption{Signals recorded by the DC pad (blue) and by the closest AC pad (red) when the impact point is in the outer/inner part of the RSD. The solid lines are a single pulse, while the average on few tens of waveforms is represented with markers.}
    \label{fig:DCpulse}
\end{figure}

\begin{itemize}
\item impact points between the edge of the sensor and the first column/row of metal AC pads: both the DC (blue) and AC (red) signals have a fast component.  
\item impact points anywhere among the AC pads: the DC signal does not show a fast component. 
\end{itemize}

This happens because the fast component always follows the path with the lowest high-frequency resistance to ground: when the particle hits among AC pads, this is represented by the AC ground. However, when the particle hits near the edge, the DC contact is sufficiently low resistance to absorb part of the fast signal. 

\begin{figure}[th!]
    \centering
    \includegraphics[scale=0.2]{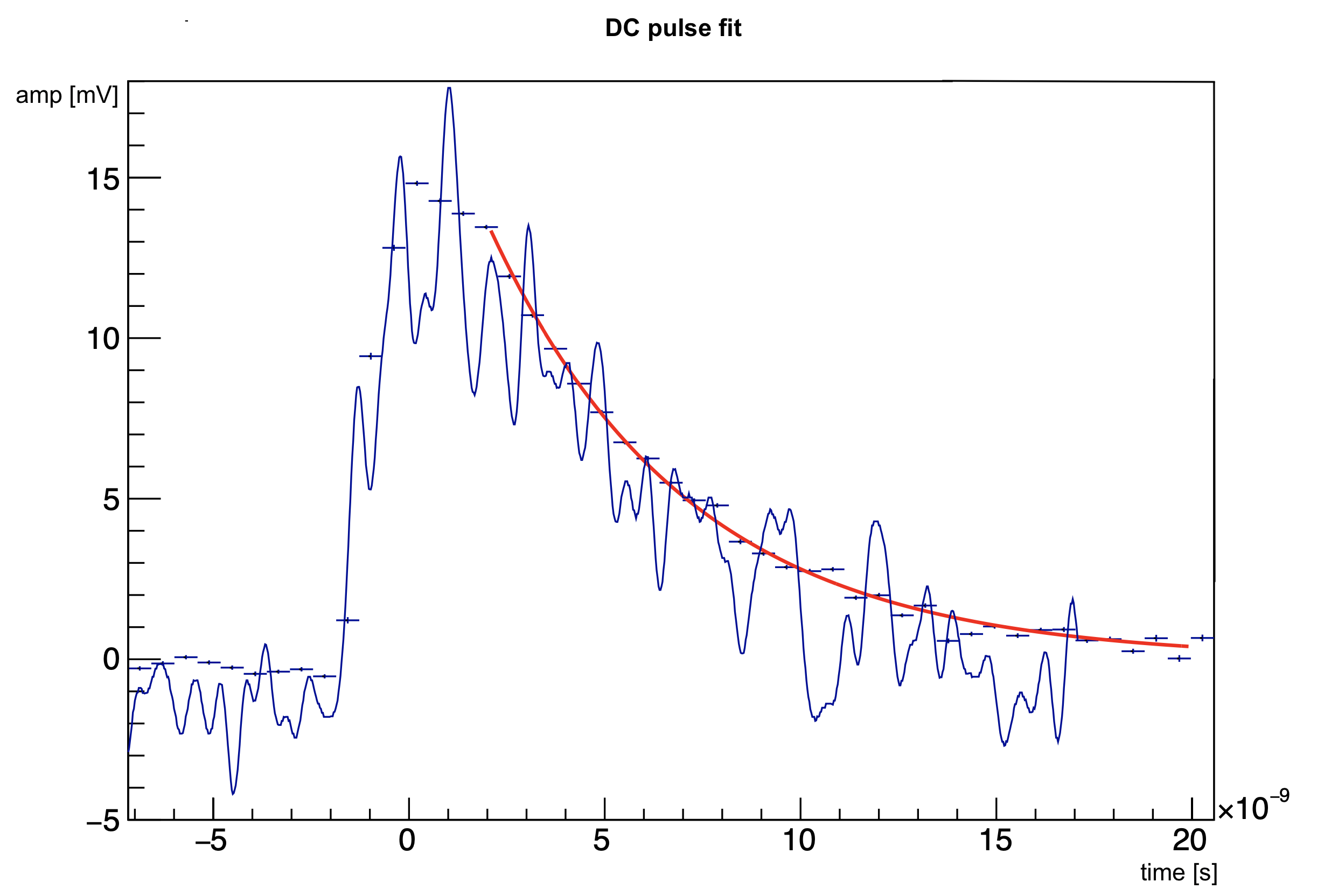}
    \caption{The average of a hundreds of DC signals (markers) selected in the DC pad inner region and fitted with an exponential function to measure the RC time constant (red line). An example of a single pulse is represented in a solid blue line.}
    \label{fig:RCtime}
\end{figure}

In both regions, the falling edge of the pulse depends upon the AC discharge current. The RC time constant of the DUT can be obtained by fitting the falling edge of DC signals with the exponential function $e^{-\frac{t}{RC}}$ 
(Figure \ref{fig:RCtime}). 
The value obtained for this geometry is $(5.08\pm0.03)\;ns$. 

\begin{figure}[th!]
    \centering
    \includegraphics[scale=0.16]{ 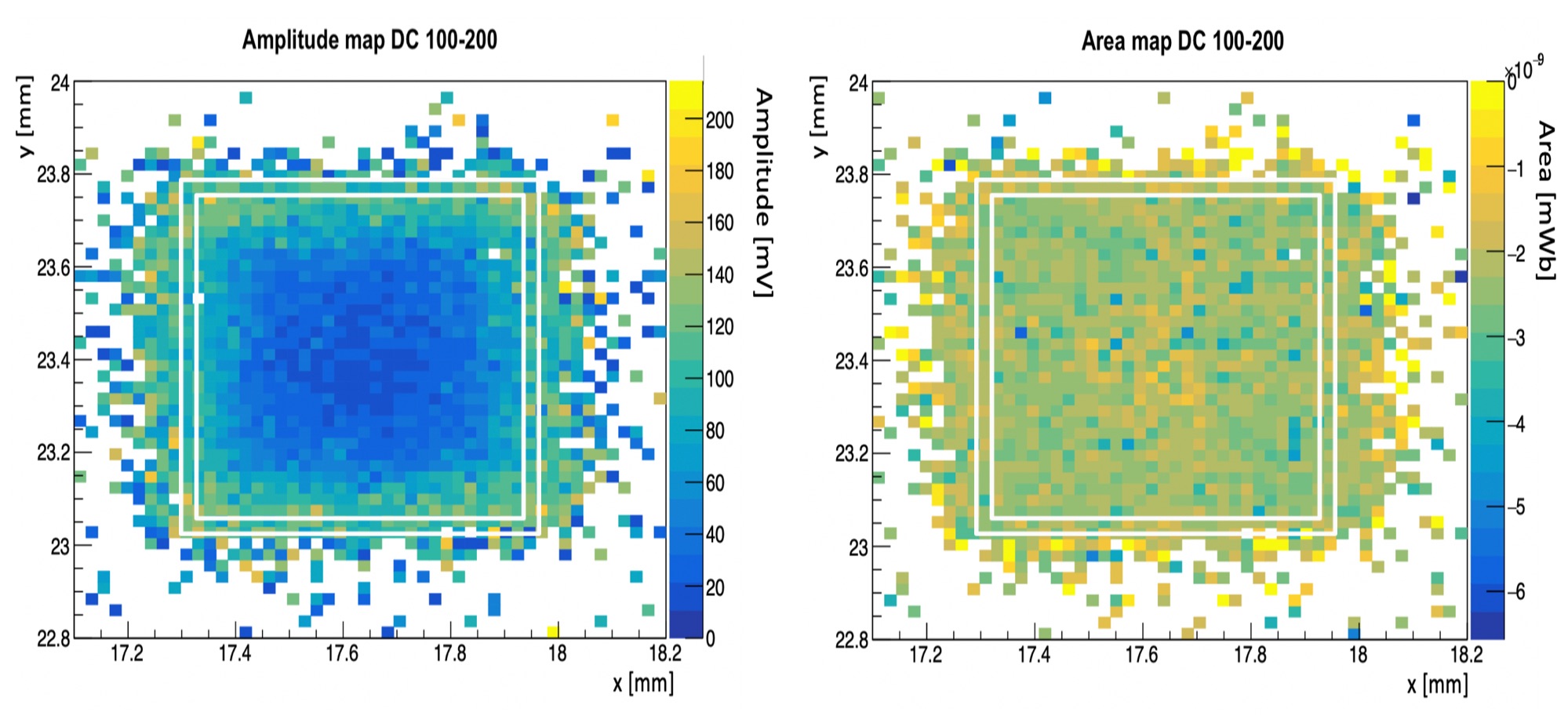}
    \caption{Left: the amplitude of the DC pad signal shows a difference between the inner and outer regions. Right: the area of the DC pad signal is constant. 
    The DC contact is outlined in white.}
    \label{fig:amp+areaDC}
\end{figure}

The amplitude of the DC signal as a function of the position is displayed in the 2D map in Figure \ref{fig:amp+areaDC} (left): pulses near the DC contact (outlined in white) have a much larger amplitude due to the presence of the fast component. On the contrary, the pulse area, shown on the right side of Figure \ref{fig:amp+areaDC}, remains constant.

Given the variability of the signal shape upon the impact point position, it is not possible to perform precise timing measurements with the DC contact.




\section{Conclusions}
This paper presents the principles of operation of Resistive AC-Coupled Silicon Detectors, a new type of LGAD characterized by a continuous gain layer and an AC read-out design. RSDs are the first silicon devices with internal signal sharing: by design, signals are spread among neighbouring pads, allowing for a very precise determination of the particle impact point position. 
The paper presents the results of extensive studies performed using a laser TCT system and beam test data, combined to evaluate the RDS performances.
Studies with the laser system point to spatial resolutions of  $\sim 5\; \mu m$  with pixels size up to ($\sim 200\; \mu m$). The temporal resolution, evaluated at beam test with protons of 120 GeV/c momenta, has been measured to be $\sim 40 \; ps$.

\section*{Acknowledgment}

We kindly acknowledge the following funding agencies and collaborations: INFN - Gruppo V; Horizon 2020, grant UFSD 669529; Dipartimenti di Eccellenza, Universit\'a degli Studi di Torino (ex L. 232/2016, art. 1, cc. 314, 337); Ministero della Ricerca, PRIN 2017, progetto 2017L2XKTJ 4DInSiDe; Ministero della Ricerca, FARE, progetto R165xr8frtfare.







\end{document}